\documentclass[twocolumn,pre,showpacs,superscriptaddress,floatfix]{revtex4}

\usepackage{calligra}
\DeclareMathAlphabet{\mathcalligra}{T1}{calligra}{m}{n}

\usepackage[normalem]{ulem}
\usepackage{xcolor}
\usepackage{graphicx}
\usepackage{bm}
\usepackage{epstopdf}
\usepackage{amsbsy,amsmath}
\usepackage{amsfonts,paralist}
\usepackage{verbatim} 
\usepackage{comment}
\newcommand{\BEQ}{\begin{equation}}
\newcommand{\EEQ}{\end{equation}}
\newcommand{\BEA}{\begin{eqnarray}}
\newcommand{\EEA}{\end{eqnarray}}
\renewcommand{\H}{{\cal {H}}}

\makeatletter
\newcommand\figcaption{\def\@captype{figure}\caption}
\makeatother

\begin{document}
\title{Small clusters Renormalization Group in 2D and 3D Ising and BEG models with
ferro, antiferro and quenched disordered magnetic interactions} \date{\today}

\author{F. Antenucci} \affiliation{IPCF-CNR,
  UOS Roma {\em Kerberos}, P.le Aldo Moro 2, I-00185 Roma, Italy}
\affiliation{Dipartimento di Fisica, Universit\`a "Sapienza", P.le
  Aldo Moro 2, I-00185 Roma, Italy}

\author{A. Crisanti} \affiliation{Dipartimento di Fisica, Universit\`a
  "Sapienza", P.le Aldo Moro 2, I-00185 Roma, Italy}
\affiliation{ISC-CNR, UOS Sapienza, P.le Aldo Moro 2, I-00185 Roma,
  Italy}

\author{L. Leuzzi} \email{luca.leuzzi@cnr.it} \affiliation{IPCF-CNR,
  UOS Roma {\em Kerberos}, P.le Aldo Moro 2, I-00185 Roma, Italy}
\affiliation{Dipartimento di Fisica, Universit\`a "Sapienza", P.le
  Aldo Moro 2, I-00185 Roma, Italy}

\begin{abstract}
The Ising and Blume-Emery-Griffiths (BEG) models critical behavior is analyzed in 2D and 3D by means of a 
renormalization group scheme on small clusters made of a few lattice cells. Different kinds of cells 
are proposed for both ordered and disordered model cases. 
In particular, cells preserving a possible antiferromagnetic ordering  under renormalization allow for 
the determination of the N\'eel critical point and its scaling indices. 
These also provide more reliable estimates of  the Curie fixed point  
than those obtained using cells preserving only the ferromagnetic ordering.
In all studied dimensions, 
the present procedure does not yield a strong disorder critical point corresponding to the
transition to the spin-glass phase. 
This limitation is thoroughly analyzed and motivated.
\end{abstract}

\maketitle

\section{Introduction}

In this work we shall discuss the real space Renormalization Group (RG) study of  critical 
behavior of spin systems interacting via  different types of magnetic interaction.
We will consider the Ising and the Blume-Emery-Griffiths (BEG) models, 
where spins can take either the value $\pm 1$, magnetic site,  or $0$,   {\em hole}.

The real space RG is based on a number of RG transformations. 
Different RG transformations have been used in literature, all sharing the property of being ``simple", i.e.,  
the space of allowed couplings must be kept low-dimensional avoiding their proliferation. 
This process necessarily involves arbitrary and uncontrolled approximations.
One possible approach is to replace the original lattice by a different lattice obtained by a bond-moving procedure.
This is the case of the hierarchical lattices 
(see Ref. [\onlinecite{Kaufman81, Kaufman82}] for the definition and,
e.g., Ref. [\onlinecite{Salmon2010}] for a recent summary of the achievements).
The main drawback is that hierarchical lattices are quite inhomogeneous and have geometrical properties that differ 
from those of Bravais lattices even locally, sometimes leading to different physical behaviors \cite{ACL_HL}.

In this work, instead, we employ alternative \textit{cell blocks transformations}, 
such as those proposed in the seventies, e.g., in
Refs. [\onlinecite{Niemeijer73,Berker76}].
In particular, this approach is proven reliable in the study of the percolation problem \cite{Percolation77, Percolation78},
where each site is present with a given probability, independent of the state of the neighbors sites.
When site interactions are introduced,
this real space RG approach has turned out to be quite powerful for studying the Paramagnetic (PM) -- Ferromagnetic (FM)
transition, while it often fails to detect more complex phases, 
such as the antiferromagnetic (AFM) phase in system with antiferromagnetic interactions 
or the spin-glass (SG) phase in disordered systems.

In this paper we start from the cluster approximation for ferromagnets used by  Berker and Wortis
\cite{Berker76} and we will consider possible generalizations to more structured block RG transformations
to capture the N\'eel point of antiferromagnetic systems, and we will analyze the robustness of both the 
FM Curie and AFM N\'eel critical points to a small amount of disorder. We shall also investigate 
the possibility of  the onset of a SG critical point in the case of strong quenched disorder.

The construction of the block RG transformation is regulated by two opposite requirements: (i) minimal 
cluster structure to capture the properties of the phases and (ii) computational feasibility.
%
In particular the last request again
results in a ``hierarchical structure'' of the system, different from the original Bravais lattice,
such to prevent the development of different kinds of interaction at every RG step.
However, in contrast to hierarchical lattices, 
the local geometry of the Bravais lattice is preserved.

We will consider the critical behavior in both 2D and 3D dimensions, and compare our results to 
the outcome of numerical simulations and, for small disorder, to the
predictions of the \textit{gauge theory} of Nishimori \cite{Nishimori01}.

The paper is organized as follows. Section \ref{s:psrgI2D} is devoted to 2D Ising models. 
Here we also recall the real space block RG transformation procedure, and its extension to the
case of (quenched) random interactions. We also introduce the generalization of the block RG 
transformation used to tackle  antiferromagnetic and disordered interactions.
 In Sec. \ref{s:Ising3D} we extend the analysis  to 
 3D Ising models
 and in Sec. \ref{s:BEG} to the Blume-Emery-Griffiths model.
 
 Finally, in Sec. \ref{s:discussion} we summarize our findings and we comment  about the inability 
 to locate a SG critical point for strong disorder, and how it might be overcome.
 
\section{Cluster Renormalization Group for the 2D Ising model}
 \label{s:psrgI2D}

 The real space block RG transformation dates back to the 70's, and consists of the following steps:
\begin{enumerate}
\item {\em group} spins on the real space Bravais lattice into blocks 
with a given geometry;
\item {\em replace} each block by a  new spin variable,  {\em block-spin}, whose value is dictated 
by the values of all the spins inside cell through a 
{\em projection matrix};
\item {\em sum} in the partition  over all spins inside the cells for fixed
value of the block-spins;
\item {\em rescale} the lattice-space to its original value and compute the new, 
{\em renormalized}, values of interactions among the block-spins leaving the partition function invariant.
\end{enumerate}

When points 1 to 4 are iterated they yield the RG flow
$\boldsymbol{\mathcal{K}}_R = \mathcal{R}(\boldsymbol{\mathcal{K}})$
in the interaction parameters space $\boldsymbol{\mathcal{K}}$. Starting from the initial physical 
values the renormalized parameters flow towards a fixed point 
$\boldsymbol{\mathcal{K}}^* = \mathcal{R}(\boldsymbol{\mathcal{K}}^*)$
that characterizes the phase of the system. The stability matrix of the fixed point  gives 
the critical exponents.

In this Section we apply this procedure to the 2D Ising model with quenched  disordered bimodal
ferromagnetic/antiferromagnetic interactions. The Hamiltonian, expressed in a form suitable for the  
RG study, is
\begin{equation}
- \beta \mathcal{H}(\bm{s}) = \sum_{\langle ij \rangle} \left[ J_{ij} s_i s_j + h_{ij} \frac{s_i+s_j}{2} + h_{ij}^{\dagger} \frac{s_i - s_j}{2} \right],
 \label{f:Ham_Ising}
\end{equation}
where $\langle ij\rangle$ denotes the ordered sum over nearest-neighbor sites on the 2D Bravais lattice. 
As usual in RG studies, we use reduced parameters where the temperature is absorbed into the 
interactions parameters. 

The initial (physical) probability distribution of the couplings is
\begin{align}
 P({\mathcal{K}}_{ij})  =&  
             P(J_{ij}) \, P(h_{ij}) \,P(h^{\dagger}_{ij})
 \nonumber\\
 =&
 \bigl[ (1-p) \delta (J_{ij} + J) + p \delta (J_{ij} - J) \bigr]   \nonumber
 \\
 & \phantom{==} \times \delta  \bigl( h_{ij} -h\bigr)\, \delta\bigl( h_{ij}^\dagger \bigr) \,  ,
\label{f:P_J}
\end{align}
where $\boldsymbol{\mathcal{K}}= \{J, h, h^{\dagger}\}$.

\subsection{Ferromagnetic 2D Ising model}
\label{sec:2DIFM}

\begin{figure}[t!]
\centering
\includegraphics[width=\columnwidth]{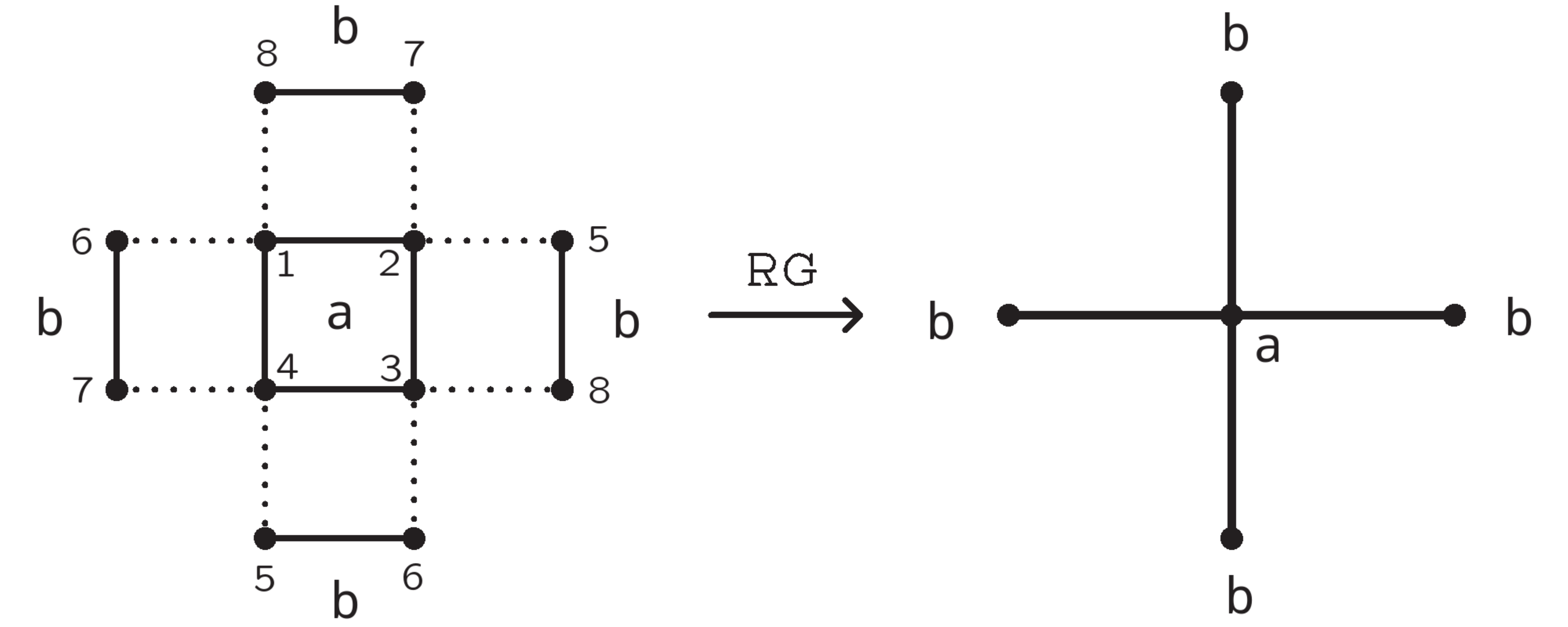}
\caption{ $\text{SQ}_2$ cluster  [\onlinecite{Berker76}]: 
                two square cells $C=a, \, b$ are arranged with periodic boundary conditions.
                Full line denotes intra-cell bonds, while dotted lines inter-cell bonds.  
                 Under the block RG transformation the cells are replaced by the block-spins 
                  $\sigma_{a,b}$.
                 The periodic boundary conditions ensures that  each block-spin is 
                 connected to the other  one by four bonds.}
\label{fig:BW}
\end{figure}

To illustrate, and fix the notation, we shall first discuss the case of the pure ferromagnetic model
($p=1$).  Following  Berker and Wortis \cite{Berker76} we consider square cells of a 2D square lattice 
and arrange them in the cluster shown in Fig. \ref{fig:BW} (we shall refer to this
geometry as ``$\text{SQ}_2$''). 
The cluster consists of only two square cells with periodic boundary conditions.
The cell $a$ contains spins $\{s_1,\, s_2,\, s_3,\, s_4\}$
and the cell $b$ spins $\{s_5,\, s_6,\, s_7,\, s_8\}$.
This geometry ensures that the block RG transformation does not introduce additional couplings,
besides $J$, $h$ and $h^{\dagger}$.
Moreover, periodic boundary conditions guarantees that each spin has exactly four 
neighboring sites, so that the system has the correct multiplicity factor for the completely aligned
configurations. 
A correct multiplicity is a necessary, but not sufficient, requirement for a correct estimation 
of the critical temperature.

Next for each cell $c$ a new block-spin $\sigma_c$ is defined, step 2, 
using a projection matrix $\mathcal{M}(\sigma_c,s_{i\in c})$ that maps each configuration of the spin of 
the cell $s_{i\in c}$  to the value of the block-spin $\sigma_c$. 
The most general projection matrix that preserves the up-down symmetry of the Ising spins 
is  
\begin{equation*}
\begin{tabular}{ccc}
$\mathcal{M}(1,s_{i\in c}) $& \quad $\mathcal{M}(-1,s_{i\in c})$\quad & \quad ${s_{i\in c}}$   \\
\hline
$1$ & $0$ & $+ + + + $ \\
$1-t$ & $t$ & $ + + + - $ \\
$1/2$ &  $1/2$ &$+ + - - $ \\
$t$ & $1-t$&  $ + - - - $ \\
$0$ & $1$ & $----$ 
\end{tabular}
\label{eq:ProjetionMatrix_Ising}
\end{equation*}
with  $\mathcal{M}(-1,-s_{i\in c})=\mathcal{M}(1,s_{i\in c})$.
The parameter $t$ is a free parameter that controls the relative 
weight of non-symmetric configurations, and its value can be tuned to fine adjust 
the outcome of the RG analysis to known results. From its definition one may expect 
$0\leq t\leq 1$, however we will see that fine tuning may lead to $t$ outside these boundaries.
For $t=0$ one recovers the 
\emph{majority rule}: the value of the block-spin  is the value of the majority 
of spins of the cell, and $\pm 1$ with equal probability in case of parity.
We will refer to the  version of the method in which $t$ is different from zero 
(fixed to correctly obtain known critical points of the model)
as \emph{tuned} two square cells lattice (``$\text{tSQ}_2$'').

The next step is done by summing in the partition sum over all possible configurations 
of the spins  of the cells $s_{i}$ for fixed block-spins $\sigma_c$. 
This leads to the renormalized Hamiltonian
$\H'(\bm{\sigma})$
\begin{align} 
e^{-\beta \H'(\bm{\sigma})}&=
\sum_{\bm{s}}\left[ \prod_{c} \mathcal{M}(\sigma_c,s_{i\in c}) \right] e^{-\beta \H(\bm{s})} .
\end{align}
for the block-spin.

The procedure must leave the partition function invariant. Therefore, the final step is 
the replacement $\sigma \to s$ and
a rescaling that changes $\H'$  back to the original form of the Hamiltonian in the new spin $s$:
\begin{align}
 -\beta \H_R(\bm{s}) = \alpha \Bigl( J_R\, s_{a} s_b +& h_R \frac{s_a+s_b}{2}
 + h^\dagger_R \frac{s_a-s_b}{2} \Bigr) \, ,
 \label{f:Ham_Ising_BW}
\end{align}
with the renormalized interactions:
\begin{align}
\label{eq:RG1a}
 J_R &= \frac{1}{4 \alpha}  \log \biggl( \frac{x_{++}x_{--}}{x_{+-}x_{-+}} \biggl) , \nonumber \\
 h_R &=  \frac{1}{2\alpha} \log \biggl( \frac{x_{++}}{x_{--}} \biggl) , \\
 h^{\dagger}_R &=  \frac{1}{2\alpha} \log \biggl( \frac{x_{+-}}{x_{-+}} \biggl) \nonumber ,
\end{align}
where 
\begin{equation}
 \label{f:Boltzmann}
   x_{\sigma_a \sigma_b} = \sum_{\bm{s}} \mathcal{M}(\sigma_a, s_{i\in a})\,
                                                                           \mathcal{M}(\sigma_b, s_{i\in b})\, e^{-\beta \H( \bm{s})} \, 
\end{equation}
are the so-called  edge Boltzmann factors.
The coefficient  $\alpha$ is the number of near-neighbor sites on the lattice, $4$ for the 2D case.

Note that if $h=0$, then $\mathcal{H}(\bm{s})= \mathcal{H}(-\bm{s})$, implying   
$x_{+-}=x_{-+}$ and $x_{++}=x_{--}$   and, eventually, $h_R=h^{\dagger}_R=0$.

Equations (\ref{eq:RG1a})-(\ref{f:Boltzmann}) define the 
the  RG flow  $\boldsymbol{\mathcal{K}}_R = \mathcal{R}(\boldsymbol{\mathcal{K}})$.
The critical exponents are obtained from the eigenvalues of the stability matrix 
$\partial \boldsymbol{\mathcal{K}}_R / \partial\boldsymbol{\mathcal{K}}$
evaluated at the fixed point $\boldsymbol{\mathcal{K}}^*$,
which can be written in terms of
\begin{align*}
 \frac{\partial x_{s_a s_b}}{\partial J} &= \sum_{\overset{\bm{s}}{\langle ij\rangle}} 
            \mathcal{M}(\sigma_a, s_{i\in a})\, \mathcal{M}(\sigma_b, s_{i\in b})\,
           s_is_j \, e^{- \beta \mathcal{H}(\bm{s})} \, , 
\\
 \frac{\partial x_{s_a s_b}}{\partial h} &= \sum_{\overset{\bm{s}}{\langle ij\rangle}} 
            \mathcal{M}(\sigma_a, s_{i\in a})\, \mathcal{M}(\sigma_b, s_{i\in b})\,
 \frac{s_i + s_j}{2} \, e^{-\beta \mathcal{H}(\bm{s})} \, , 
\\
 \frac{\partial x_{s_a s_b}}{\partial h^{\dagger}} &= \sum_{\overset{\bm{s}}{\langle ij\rangle}}
            \mathcal{M}(\sigma_a, s_{i\in a})\, \mathcal{M}(\sigma_b, s_{i\in b})\,
 \frac{s_i - s_j}{2} \, e^{-\beta \mathcal{H}(\bm{s})}.
 \label{eq_der_BF}
\end{align*}
The nontrivial fixed point(s)  are for  $h=h^{\dagger}=0$.
In this case  the stability matrix is diagonal and 
the relevant scaling exponent are  $y_T = \log_b (\partial/\partial_J) J_R$ and 
$y_h = \log_b (\partial/\partial_h) h_R$, where $b$ is the lattice scaling factor, 
equal to $2$ for the $\text{SQ}_2$ cluster of Fig. {\ref{fig:BW}}. 
The critical exponents are then 
\begin{align}
 \nu &= \frac{1}{y_T}
\, , &
 \eta &= d+2-2y_h.
\end{align}
The others follow  from the scaling laws.

The numerical implementation of this procedure gives for the ordered ferromagnetic 2D Ising 
model ($p=1$) the critical temperature $T_{\text{c}}= J_c^{-1} =1.896$ for the PM/FM transition, and
scaling exponents $y_T=0.727 $ and $ y_h=1.942 $, see also Ref. [\onlinecite{Berker76}].
The value $y_h $ is less than the dimension of the space, implying that the transition is of the second order
 \cite{Nienhuis75}. The values of the critical exponents are shown in the first row of Table 
 \ref{tab:2d_2celle_BW}. 

By comparing with the exact Onsager solution \cite{Onsager43}, the critical temperature
deviates of about $20\%$ from the exact result
$T_{\text{c}}^{\rm Ons}=2/\log(1+\sqrt{2}) = 2.2692...$ and the values of the critical exponent
all suffer major deviations.
We postpone the discussion on how this estimates could be improved.

\begin{table}[t!]
\begin{center}
\begin{tabular}{lcccccc}
\hline
\hline
           & $\boldsymbol \alpha$  & $\boldsymbol \beta$ & $\boldsymbol \gamma$ & $\boldsymbol \delta$ & $\boldsymbol \nu$ & $\boldsymbol \eta$  \\
\hline
$\text{SQ}_2$ & -0.7523   & 0.08038  & 2.592    & 33.24    & 1.376 & 0.1168  \\
$\text{tSQ}_2$ Ons. & -0.1233   & 0.1383  & 1.847   & 14.35   & 1.062 & 0.2606  \\ 
$\text{tSQ}_2$ Nish. ~~ & -1.426   & 0.05884  & 3.309   & 57.23   & 1.713 & 0.06870  \\  
$\text{SSQ}_2$ & -0.6545   & 0.2141  & 2.226    & 11.40    & 1.327 & 0.3226  \\
$\text{SQ}_4$ & -0.1524   & 0.1915  & 1.769    & 10.24    & 1.076 & 0.3559  \\
$\text{SSQ}_4$  & -0.4458   & 0.4779  & 1.490    & 4.118    & 1.222 & 0.7815  \\
Exact     & 0	       &  0.125  & 1.75     & 15       & 1     &  0.25   \\
\hline
\hline
\end{tabular}
\caption{Critical exponents of the ferromagnetic 2D Ising Model obtained with the 
         different clusters discussed in this work compared with the known exact results.
         In the second (third) line the parameter $t$ is fixed to reproduce the known Onsager (Nishimori) critical temperature
         of the 2D lattice.
}
\label{tab:2d_2celle_BW}
\end{center}
\end{table}

\subsection{Disordered 2D Ising Model}
\label{s:quenched_RG}
In presence of quenched disorder the RG flow cannot be restricted to single interaction values
$\boldsymbol{\mathcal{K}}$, and necessarily involves the 
whole coupling probability distribution $P(\boldsymbol{\mathcal{K}})$.
The RG equation then becomes
\begin{align}
P_R(\boldsymbol{\mathcal{K}}_R)=\int d \boldsymbol{\mathcal{K}}\,
P(\boldsymbol{\mathcal{K}})\,
\delta\bigl[\boldsymbol{\mathcal{K}}_R -{\cal R}(\boldsymbol{\mathcal{K}})
\bigr].
\end{align}
The block RG transformation must then be repeated starting from 
interaction parameters configurations $\boldsymbol{\mathcal{K}}$ extracted with probability
$P(\boldsymbol{\mathcal{K}})$. The outcomes $\boldsymbol{\mathcal{K}}_R$ are then used to 
construct the renormalized probability distribution $P_R(\boldsymbol{\mathcal{K}})$, which, in turn, is
used as entry for the next iteration. 

In a numerical study, the number of possible interaction parameter configurations  that can be
considered is finite. The flow of the renormalized probability distribution 
$P_R(\boldsymbol{\mathcal{K}})$ can then be followed  by using a method initially suggested 
in Ref. [\onlinecite{Southern77}]. 
One first sets up a starting pool of $M \gg 1$ different randomly chosen real numbers 
produced according to the initial probability of the couplings, Eq. (\ref{f:P_J}) for the bimodal 
Ising Model. 
Then a coupling configuration $\boldsymbol{\mathcal{K}}$ is constructed by randomly
picking
 numbers from the pool and assigning them to the couplings. A
renormalized $\boldsymbol{\mathcal{K}}_R$ is, thus, evaluated. 
The procedure is repeated $M$ times obtaining a new pool that represents the 
renormalized probability distribution, from which one can compute the moments and
estimate $P_R(\boldsymbol{\mathcal{K}})$ from the frequency 
histogram.

\begin{figure}[t!]
\centering
\includegraphics[width=\columnwidth]{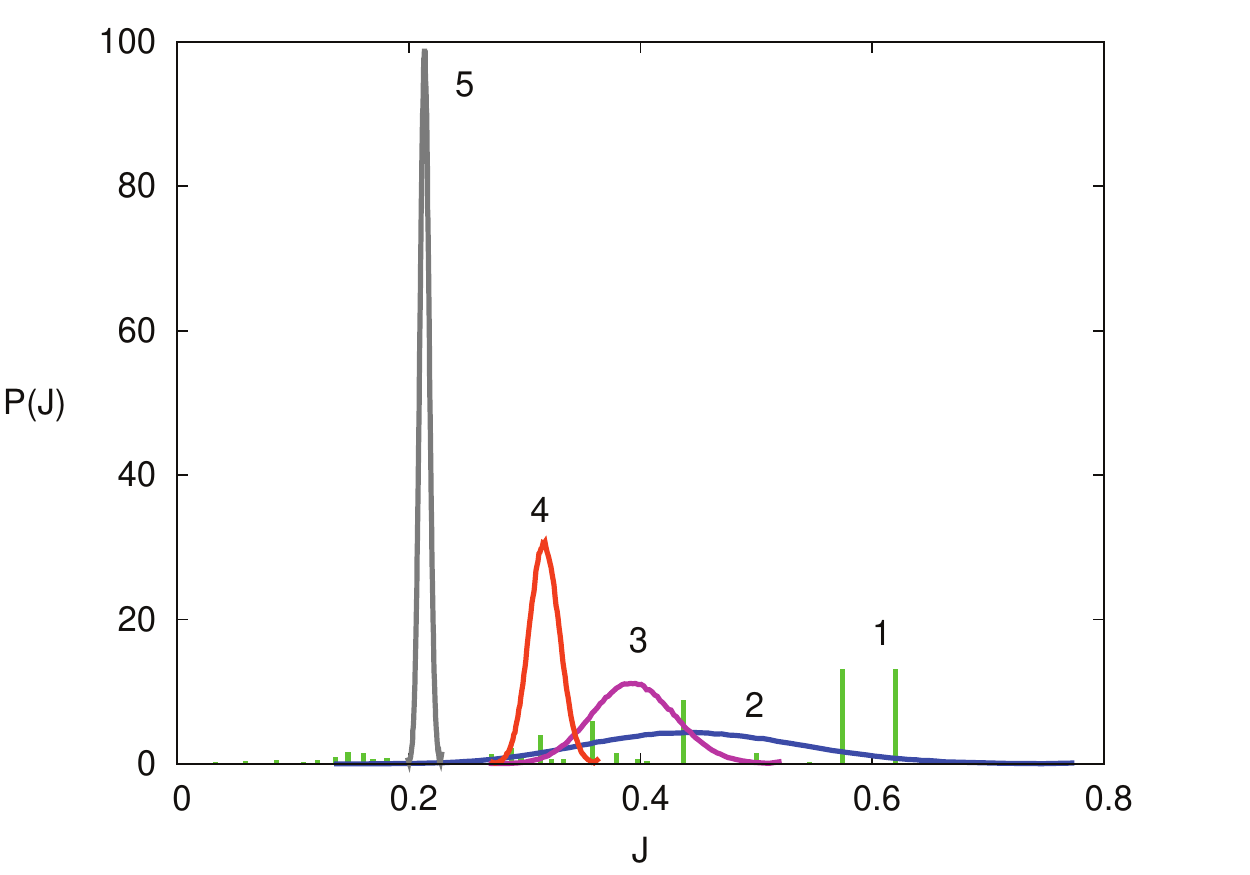}
\includegraphics[width=\columnwidth]{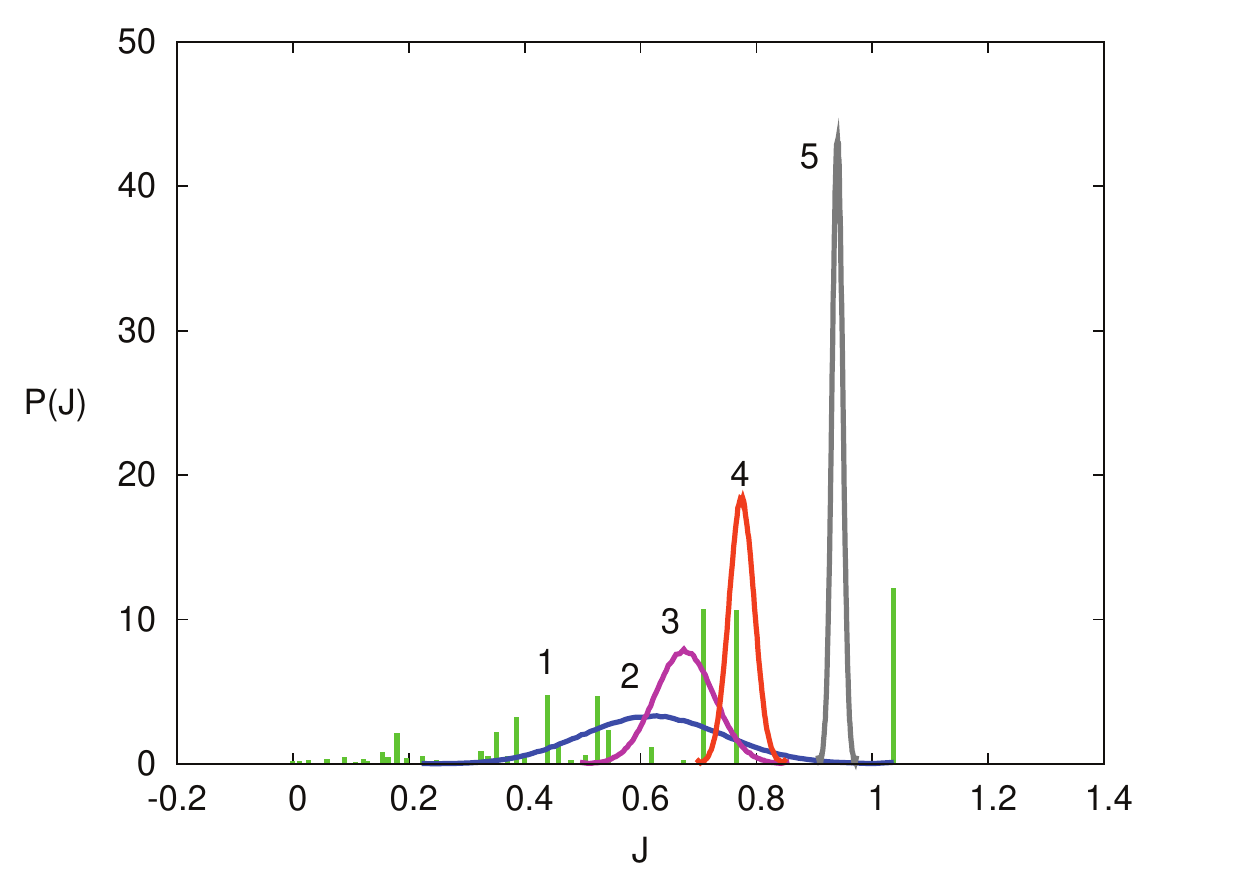}
\caption{       RG flow of the probability distribution $P(J_{ij})$ 
                for the disordered 2D Ising model (\ref{f:Ham_Ising})-(\ref{f:P_J}) with $h=0$ and $p=0.9$.
                obtained with the $\text{SQ}_2$ cluster.
                The histograms are obtained by taking $5 \cdot 10^3$ bins around the mean value 
                $\mu_J$.  The bin width is fixed by the requirement that $99.9\%$ of the values are 
                inside the grid used to follow the histogram flow.
               Upper Figure: $T=J^{-1}=1.4$, evidence for PM phase. 
               Lower Figure: $T=J^{-1}=1.2$, $p=0.9$, evidence for FM phase. }
\label{fig:RG_evolution}
\end{figure}

In Fig. \ref{fig:RG_evolution} we show the flow of the probability distribution
$P(J_{ij})$ of a single pool in the disordered 2D Ising model (\ref{f:Ham_Ising})-(\ref{f:P_J}) with $h=0$ and $p=0.9$
generated by the block RG transformation on the $\text{SQ}_2$ cluster. 
In the upper figure, $\, T=J^{-1}=1.4$, the average $\mu_J$ moves towards smaller values while the width of the distribution shrinks. 
This signals a PM phase, with the PM fixed point probability distribution function of mean $\mu_J \to 0$ and variance 
$\sigma_J^2\to 0$. 
In the lower figure, $T=J^{-1}=1.2$, the probability distribution width narrows while shifting 
towards larger value of $\mu_J$. This denotes a FM phase, with the FM fixed point probability
specified by  $\mu_J \to\infty$ and $ \sigma_J / \mu_J \to 0$.
We observe that a SG phase would be signaled by a fixed point probability distribution with
$ \sigma_J \to \infty$ while $ \mu_J /\sigma_J \to 0$, so that spins at great distance 
are still strong interacting but the sign of the interaction is not defined.

To reduce the possible bias introduced by the choice of the initial pool, Nobre et al. \cite{Nobre01} 
have proposed to repeat the block RG transformations using a set of $N_s$ samples with different initial pools of size $M$.
When close to a critical point flows originating from different pools may flow towards different fixed 
point distributions. The size of the region where the phase is not uniquely identified 
gives  the uncertainty on the critical value obtained with a the pools of size $M$.

In our numerical study of the disordered 2D Ising model we have used $N_s=20$ pools of size
$M=10^6$ each, and we have assumed a phase {\em uniquely} defined if  at least $80\%$ of 
the RG flows flow towards the same fixed distribution. With this choice the uncertainty 
is generally less than $0.1 \%$ and the systematic error considerably decreased.

\begin{figure}[t!]
\centering
\includegraphics[width=\columnwidth]{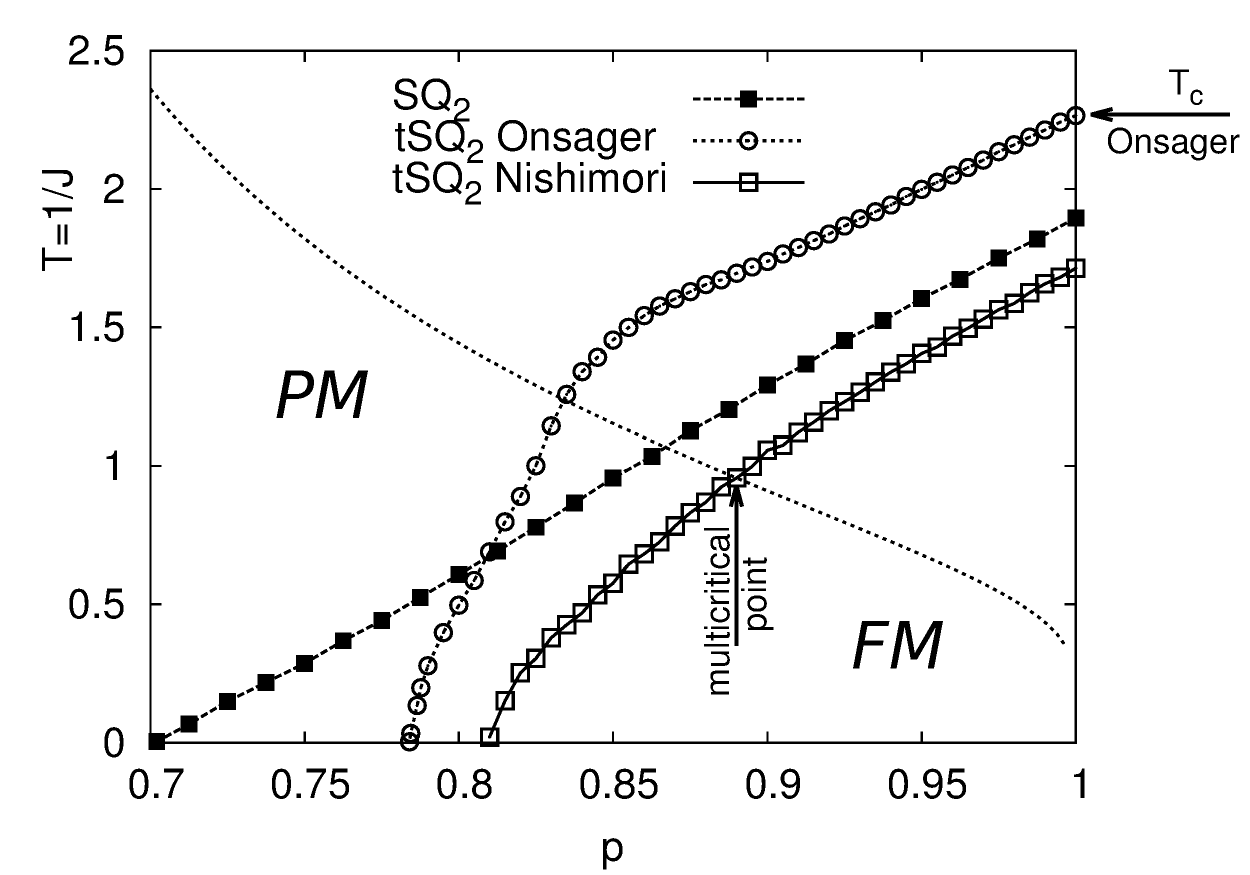}
\caption{($p,T)$ phase diagram of the disordered 2D Ising model obtained with the $\text{tSQ}_2$
                cluster and different choices for the parameter $t$ in the projection matrix.
                Filled square: $t=0$; 
                Empty circle:   $t=-0.06453$ fixed by the requirement $T_{\text{c}}(p=1) = T_{\text{c}}^{\rm Ons}$ (Onsager);
                Empty square: $t=0.0304$  fixed by the requirement $T_{\text{c}}(p_{\text{mc}}) = T_{\text{mc}}$ (Nishimori);
                Dashed line: Nishimori line.
                }
\label{fig:PhDi_2D_BWvneq0}
\end{figure}

The $(p,T)$ phase diagram of the disordered 2D Ising model obtained using the $\text{SQ}_2$ cluster
is shown in  Fig. \ref{fig:PhDi_2D_BWvneq0} (black squares).
As the probability $p$ of the ferromagnetic bonds is lowered
the critical temperature decreases until, for low enough $p$, the FM phase disappears.
In the figure also the Nishimori line \cite{Nishimori81}
\BEQ \frac{1}{T} = \frac{1}{2}\log \frac{p}{1-p}
\label{f:nishimori}
\EEQ 
is shown. Along this line the model is invariant under the \textit{gauge transformation} 
of spins and interactions and exact information about the phase diagram
can be obtained \cite{Nishimori01}.
The point where the Nishimori line crosses the transition line is called ``multicritical'': 
when a SG phase is actually present, this is the
point at which PM, FM and SG phases  all are in contact with each other.

By the RG on the $\text{SQ}_2$ cluster the 
``multicritical'' point is found at $p_{\text{mc}}=0.8667$, $T_{\text{mc}}=1.070$. 
For $p<p_{\text{mc}}$, exact results impose no FM ordering \cite{Nishimori01}.
Inspection of the Figure shows that not only the method fails to predict the correct critical temperature
$T_{\text{c}}^{\rm Ons}$ 
of the pure ferromagnetic model, but also the requirements following from the gauge theory.

One can try to improve the numerical estimates tuning the parameter $t$ in the projection matrix 
to fix some known points in the $(p,T)$ diagram. 
We consider two possible choices: fixing the critical temperature of the pure system to the exact value 
or the crossing point with the Nishimori line to the multicritical point. 
The requirement $T_{\text{c}} = T_{\text{c}}^{\rm Ons}$ leads to $t=-0.06453$, while
the requirement $T_{\text{c}}(p_{\text{mc}}) = T_{\text{mc}}$ to $t=0.0304$.
Note the ``unphysical'' negative value of $t$, also used by Berker and Wortis \cite{Berker76},  which implies that
under the block transformation the contribution of some spin configurations of the cell
to the partition sum can be negative.
The transition lines obtained with these choices for $t$ are shown in 
Fig. \ref{fig:PhDi_2D_BWvneq0}. 
In both cases, and besides the unphysical values of $t$,  
the slope of the transition line increase as $p$ decreases, but still no re-entrance or vertical line is recovered.
In either cases the only critical point remains the FM fixed point at $p=1$ with scaling exponents 
$y_T=0.9419$ and $y_h = 1.870$ for $t=-0.06453$, and  $y_T= 0.5837$ and $y_h = 1.965$ for 
$t=0.0304$. The numerical values of the critical exponents are shown in the second and third row
of Table \ref{tab:2d_2celle_BW}, respectively. Note that  in all cases  $\alpha<0$. 
According to the Harris criterion \cite{Harris74} this indicates that 
the FM fixed point is stable against the
introduction of a small amount of quenched disorder.

Summarizing the results: the block RG transformation based on the $\text{SQ}_2$ cluster 
finds no true multicritical point, nor a ``strong disorder'' fixed point,
and, hence, no change in the universality class of the critical behavior is detected.

\subsection{Antiferromagnetic order: need for ``$\text{SSQ}_2$''}
\label{s:Ising2D_snake} 
Another important issue of the block RG transformation discussed so far
is the absence of an AFM phase. Below some critical value of $p$
and down to $p=0$,  only the PM phase is found. 
This failure might also strongly bias the
quest for a spin-glass phase in dimension higher than two.  

By analyzing the block RG transformation used so far, we see that it assigns 
the same weight to symmetric configurations (e.g.,  $++--$) of the spins of the cell, regardless of
their ordering. As a consequence, it is not able to identify an antiferromagnetic ordering, and 
a staggered magnetization cannot be properly defined. 

\begin{figure}[t!]
\centering
\includegraphics[width=\columnwidth]{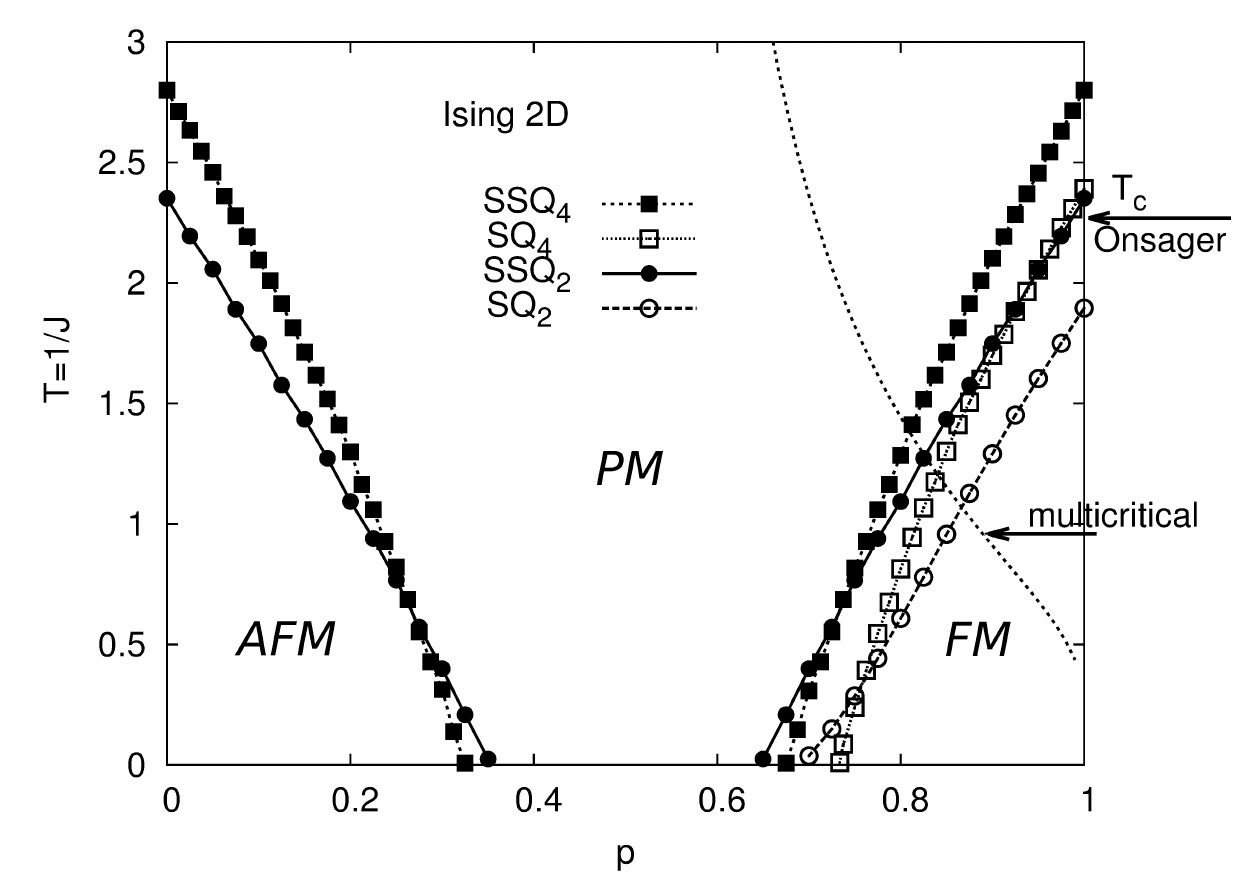}
\caption{$T$, $p$ phase diagram of the Ising 2D model as obtained by
               iterating the RG on different clusters with two and four cells.
                The dashed line is Nishimori line, Eq. \ref{f:nishimori}. 
}
\label{fig:Ising2D_Tp_PhDi}
\end{figure}

We thus need a cluster construction that distinguishes the symmetry breaking ordering 
associated with the AFM phase. By referring to labeling of Fig. \ref{fig:BW}, we then assign the spins 
$\{s_1,s_3,s_5,s_7\}$  to the cell $a$ and  the spins $\{s_2,s_4,s_6,s_8\}$ to the cell $b$, 
shaping a \emph{staggered topology} (``$\text{SSQ}_2$'' in the following).
The projection matrix of the cell remains unchanged.
The phase diagram obtained through this block RG transformation is shown in 
Fig. \ref{fig:Ising2D_Tp_PhDi}.  The improvement with respect to the $\text{SQ}_2$ cluster  is evident. 
The $p=0$  antiferromagnetic critical point is now found, as well as a PM/AFM transition line for $p>0$.
Since in this model, for $h=0$  the symmetry $(p,J)\leftrightarrow(1-p,-J)$ holds 
and the staggered cluster  preserves AFM ordering,  the PM/AFM line is symmetric to the 
PM/FM line with respect to $p=1/2$.
The behavior of the critical line below $p_{\text{mc}}$, however, still
violates the requirement imposed by the gauge theory.

The $\text{SSQ}_2$  cluster improves the estimate of the pure critical fixed point ($p=1$).
The critical temperature turns out $T_{\text{c}}(p=1)=2.352$ and deviates of about $3.5\%$ 
from Onsager result.  
The scaling exponents are $y_T =0.7534 $ and $ y_h=1.839$, and the 
associated critical exponents are reported in Table \ref{tab:2d_2celle_BW}. 
Though they display differences of $20\%$ to $40\%$  
from the exact values, their estimates are sensitively better than those obtained with the classic 
$\text{SQ}_2$ cluster.

As the AFM transition is concerned, the behavior is specular to that of  the FM
transition. The points along the AFM critical line are attracted by a unique second order 
AFM N\'eel fixed point at $p=0$ at the same critical temperature $T_{\text{c}}(p=0)=2.352$  with 
scaling exponents $y_T=0.7534$, as found for the FM fixed point, and $y_h=0.01565$.
The symmetry of the RG equations implies that the PM/AFM and PM/FM fixed 
points have the same $y_T$. The values of $y_h$ are, however, quite different, the AFM one being 
 almost zero. The reason is that the magnetization is not the correct order parameter for the 
AFM transition, as it remains	zero on both sides of the transition. 
If, rather, the {\em staggered} magnetization is considered, and, hence, a {\em staggered} field $h^\dagger$ is introduced
in the  Hamiltonian, then the relevant scaling exponent turns out to be
$y_{h^\dagger} = \log_2 \partial_{h^\dagger} h^\dagger_R = 1.797 \gg y_h$.

\subsection{$4$-square cells cluster (``$\text{SQ}_4$'')}
\label{ss:4c}

The $\text{SSQ}_2$ cluster leads to an AFM fixed point, and improves both the analysis of 
the AFM and FM phases. However, it does not allow for possible frustrated configurations in the 
renormalized cells.
In an attempt to circumvent this problem we extend the cluster from two to four cells.

\begin{figure}[t!] 
\centering
\includegraphics[width=\columnwidth]{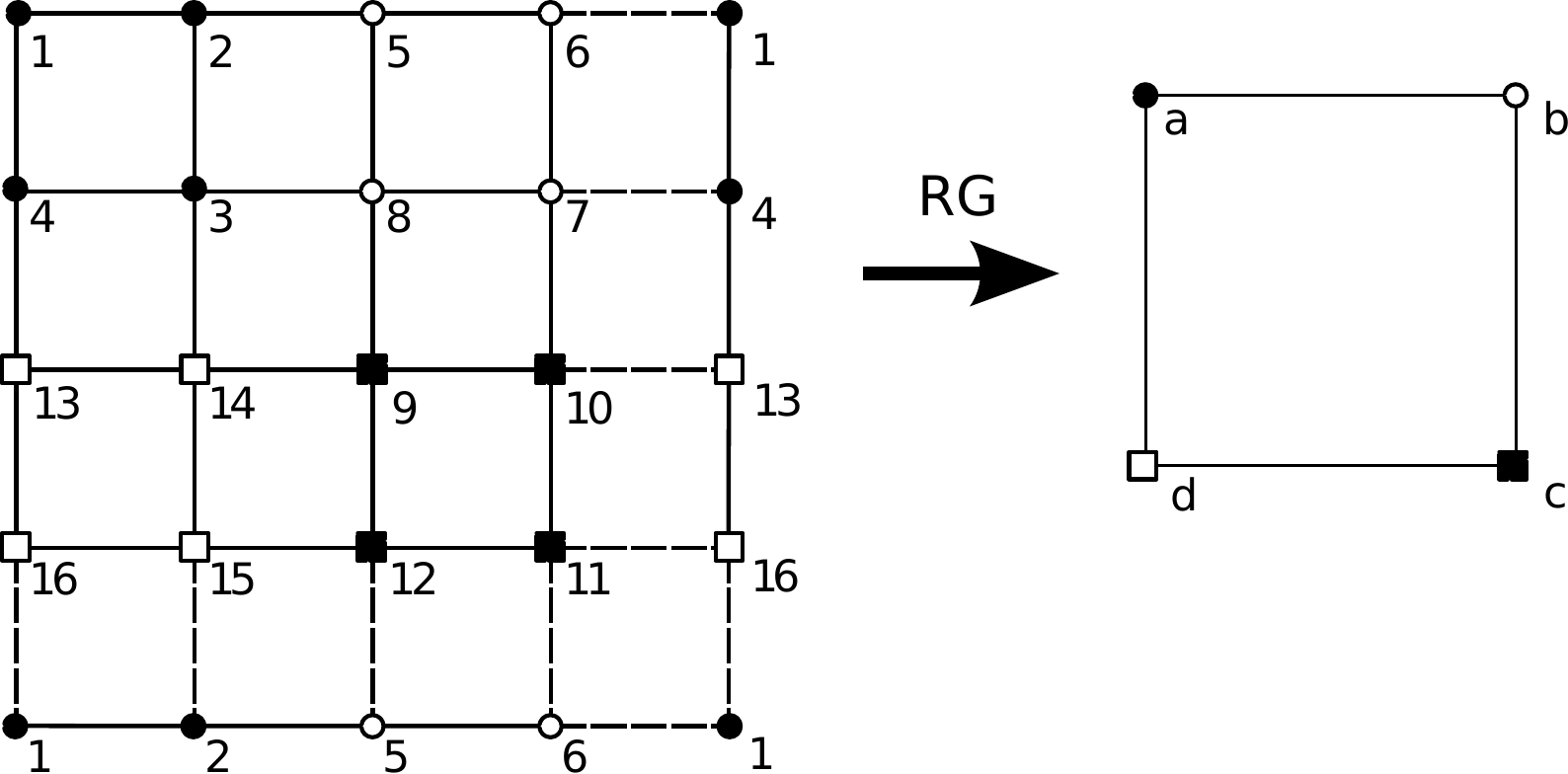}
\caption{The $16$-spins $\text{SQ}_4$ cluster: before (lhs.) and after (rhs.) renormalization.  
                Full lines denote intra-cell bonds, while dotted lines inter-cell bonds due to
                periodic boundary conditions.
                Block-spins $\sigma_{a,b,c,d}$ on the rhs. cluster are constructed from spins 
                $s_{i=1,\ldots,16}$ denoted 
                by the same symbol on the lhs. cluster. 
               }
\label{4celle_2d}
\end{figure}

As the number of cells increases, so does the number of possible cluster definitions. 
We found that the best block RG transformation, 
in terms of similarity with the exact results, is
obtained with the cluster shown in Fig.  \ref{4celle_2d}. 
The block RG transformation is 
performed by summing in the partition sum over all possible configurations 
of the spins  of the cells $s_{i=1,\ldots, 16}$ for fixed block-spins
\begin{align*}
  \sigma_a &:=  \{ s_1, s_2, s_3, s_{4} \} \, ,\\
  \sigma_b &:=  \{ s_5, s_6, s_7, s_{8} \} \, , \\
  \sigma_c &:= \{ s_9, s_{10}, s_{11}, s_{12} \} \, ,\\
  \sigma_d &:= \{ s_{13}, s_{14}, s_{15}, s_{16} \} \, . 
\end{align*}
The $16$ spins of the SQ$_4$ cluster, together with the inter-cell interactions  from periodic 
boundary conditions, form a $4\times 4$ array of $4$-spin cells.
The block RG transformation generates, besides nearest-neighbor interactions, 
also next-nearest-neighbor interactions and ``plaquette'' interactions. 

To avoid truncations we, 
then, start from the more general Hamiltonian
\begin{align} 
\label{eq:HamSQ4}
 - \beta \mathcal{H}(\bm{s})  = &  \frac{1}{2}  \sum_{{\bm i}} \sum_{k=1}^4J_{{\bm i}, {\bm i+\boldsymbol \mu_k}} s_{\bm i} s_{\bm i+\boldsymbol \mu_k} \, +  \nonumber \\
&  +  \frac{1}{2}  \sum_{{\bm i}}\sum_{k=1}^4 K_{\bm i,{\bm i+\boldsymbol \eta_k}} s_{\bm i} s_{\bm i+\boldsymbol \eta_k} \, +\\
& +  \sum_{{\bm i}} D_{\bm i} \prod_{k=1}^4 s_{\bm i+\boldsymbol \xi_k} \, ,\nonumber
\end{align}
where ${\bm i}=(i_x,i_y)$ denotes a site on the 2D lattice,
$\boldsymbol \mu$ the relative position of the nearest-neighbor sites, 
$\boldsymbol \eta$ the relative position of  the next nearest-neighbor sites
and $\boldsymbol \xi$ the relative position of the plaquette sites:

\begin{tabular}{llll}
{\phantom a}\\
\hspace*{-.3cm}$\boldsymbol \mu_1=(0,1)\, ,$  & $\boldsymbol \mu_2=(1,0)\, ,$ &
$\boldsymbol \mu_3=(0,-1)\, ,$ & $\boldsymbol \mu_4=(-1,0)\, ,$ \\
{\phantom a}\\
\hspace*{-.3cm}$\boldsymbol  \eta_1=(1,1)\, ,$   & $\boldsymbol \eta_2=(1,-1)\, ,$ &
$\boldsymbol \eta_3=(-1,-1)\, ,$ & $\boldsymbol \eta_4=(-1,1)\, ,$ \\
{\phantom a}\\
\hspace*{-.3cm} $\boldsymbol  \xi_1=(0,0)\, ,$  & $\boldsymbol \xi_2=(0,1)\, ,$ &
 $\boldsymbol \xi_3=(1,1)\, ,$   & $\boldsymbol \xi_4=(1,0).$ \\
{\phantom a}
 \end{tabular}
The initial distributions of the couplings  is
\begin{align*}
  P(\boldsymbol{\mathcal{K}}_{ij}) \!\! = \! \left[ (1\!-p) \delta (J_{ij} +\! J) + p \delta (J_{ij} -\! J) \right] 
  \delta \! \left( K_{ij} \right) \delta \! \left( D_{i} \right) \, .
\end{align*}

\begin{figure}[t!]
\includegraphics[width=\columnwidth]{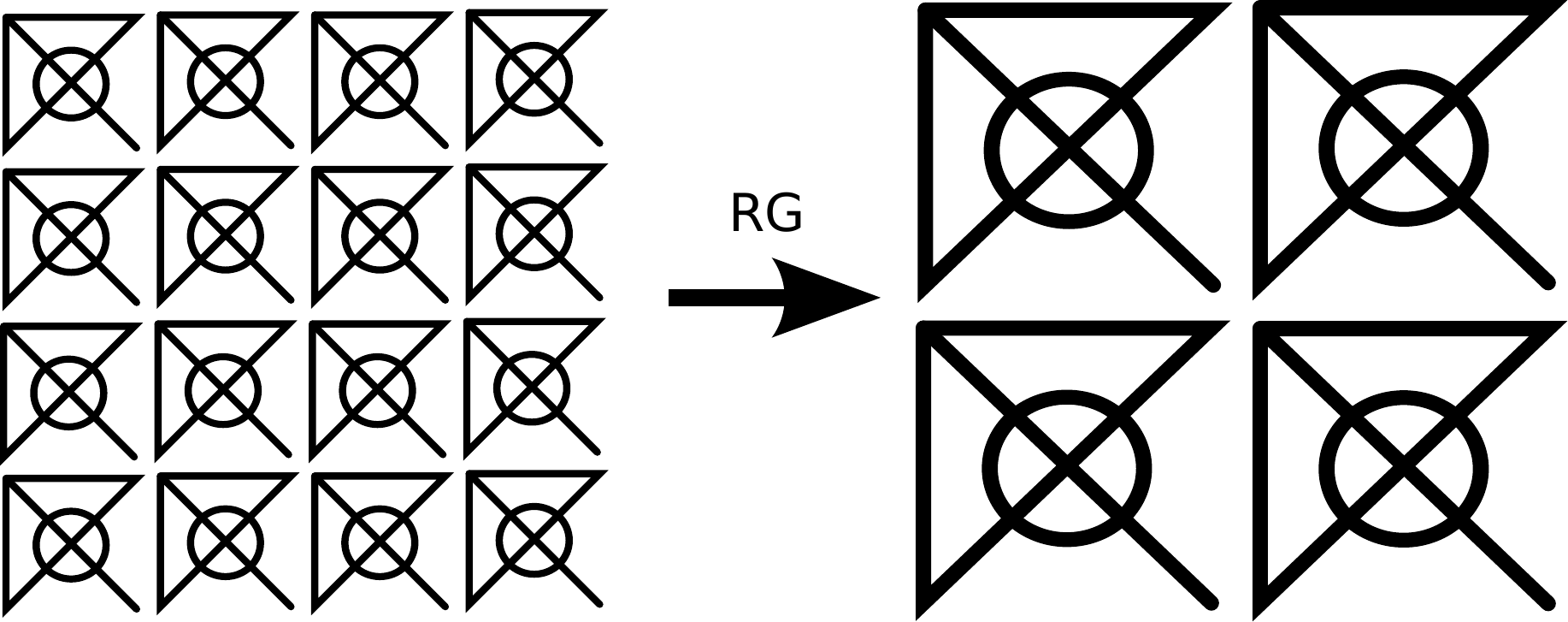}
\caption{
Allocation of the interactions for the SQ$_4$ cluster.
The interactions $J,$ $K,$ $D$ are represented respectively by horizontal and vertical lines, diagonal lines, and circles.
Ensembles of adjacent 2$J$s, 2$K$s and 1 $D$ are grouped together to form ``arrow packages'' 
that completely cover both the initial and the renormalized clusters of Fig. \ref{4celle_2d}.
These ensembles are the building blocks of the RG procedure, so that the correlation between 
the interactions in the ensembles 
is preserved under the renormalization.
}
\label{schema_4celle_2d}
\end{figure}

To best preserve the correlation between the interactions $J$, $K$ and $D$ 
we build the pools that numerically represent the interaction probability distribution by correlated ensembles 
consisting of 2 $J$s, 2 $K$s and 1 $D$ adjacent to each other.
These are the maximum sets that completely cover both the initial and the renormalized clusters, as shown in Fig. \ref{schema_4celle_2d}.
Notice that, because of the periodic boundary conditions, the four ensembles in the renormalized system of Fig. \ref{schema_4celle_2d}
only differ in the values for the $J$s, while the $K$s and the $D$ are always the same.

The block RG  with the Hamiltonian (\ref{eq:HamSQ4}) leads to four $J$-like ($J_R^{\pm}  , \, \tilde{J}_R^{\pm} $), 
two $K$-like ($K_R^{\pm}$) and one $D$-like ($D_R$) renormalized interactions whose values are
\begin{align}
 J_R^{\pm} &= \frac{1}{16}   \left( \log \frac{x_{++++} \, x_{--++}}{x_{+-+-} \, x_{+--+}} \pm \log \frac{   x_{++-+} \, x_{+++-}   }{  x_{+-++} \, x_{-+++}  } \right) \nonumber \\
 \tilde{J}_R^{\pm} &= \frac{1}{16}   \left( \log \frac{x_{++++} \, x_{+--+}}{x_{+-+-} \, x_{--++}} \pm \log \frac{  x_{-+++}  \, x_{+++-}   }{  x_{++-+} \, x_{+-++}    } \right)  \nonumber \\
 K_R^{\pm} &= \frac{1}{32}   \left( \log \frac{x_{++++} \, x_{+-+-} }{ x_{--++} \, x_{+--+}} \pm \log \frac{  x_{+-++} \,  x_{+++-}  }{   x_{++-+} \, x_{-+++}  } \right)  \nonumber \\
 D_R &=       \frac{1}{32}          \log \frac{x_{++++} \, x_{+-+-} \, x_{--++} \, x_{+--+}}{x_{+-++} \,  x_{+++-} \, x_{++-+} \, x_{-+++} }   ,  
\label{eq:ren4cells}
\end{align}
with the edge Boltzmann factors
\begin{align*}
 x_{\sigma_a \sigma_b \sigma_c \sigma_d} = \sum_{\bm{s}} & \mathcal{M}_a \mathcal{M}_b \mathcal{M}_c \mathcal{M}_d\,
        e^{-\beta \H(\bm{s})}, 
\end{align*}
where $\mathcal{M}_{x} \equiv \mathcal{M}(\sigma_x ,s_{i\in x})$ are the cell projection matrices.
The renormalized $J$s are assigned to the the 4 renormalized ensembles as
$\{ J_R^{+}, \, \tilde{J}_R^{-} \}$, 
$\{ \tilde{J}_R^{-} , \, \tilde{J}_R^{+}  \}$, 
$\{  \tilde{J}_R^{+} , \, J_R^{-} \}$, 
$\{ J_R^{-}, \, J_R^{+} \}$.

The phase diagram obtained with $N_s=10$ pools of size $M=10^6$  is shown 
in Fig.  \ref{fig:Ising2D_Tp_PhDi}, line $\text{SQ}_4$. 
All the points on the critical line are attracted by the pure fixed point at $p=1$
and critical temperature of $T_{\text{c}}(p=1)= 2.391$, 
about $5\%$ off  the exact 2D result. 

To evaluate the critical exponents we have to include in the
Hamiltonian an external magnetic field, and hence consider also the
three spin interaction $\sum_{\bm{i}} \sum_{k=1}^4 s_{\mathbf{i}} \,
s_{\mathbf{i}+\boldsymbol \mu_k} \, s_{\mathbf{i} + \boldsymbol
  \mu_{k+1}}$ generated by the RG.  This gives a total of five
parameters. At the pure fixed point only two are relevant with scaling
exponent $y_T =0.9292 $ and $ y_h=1.822 $.  The values of the
associated critical exponents are reported in the fifth line of Table
\ref{tab:2d_2celle_BW}

The re-entrance of the critical line below the multicritical point
$T_{\text{c}}(p)<T_{\text{mc}}$ is still absent.  However, the line
appears steeper than those obtained with the previous block RG
transformations, approaching the expected behavior of the model.
Despite this qualitative improvement, the intersection between the
transition line and the Nishimori line occurs sensitively above the
exact multicritical point, cfr. Table \ref{tab:mc_point2D}, and, as in
the previous cases, it does not correspond to a real multicritical
point..

The RG analysis indeed does not show critical fixed points besides the
pure critical point at $p=1$.  The so called {\em strong disorder}
fixed point \cite{Toldin09} is missing and the crossing is not
associated with flows towards the FM and strong disorder fixed points.

\begin{table}[t!]
\begin{center}
\begin{tabular}{lccc}
\hline
\hline
& \bfseries T$_{\text{Ons}}$ & \bfseries p$_{\text{mc}}$ & \bfseries T$_{\text{mc}}$ \\
\hline 
 $\text{SQ}_2$ & 1.896  &  $0.867$ &$1.070   $      \\
 $\text{tSQ}_2$ Ons. & 2.269 &   $0.834$    &  $1.242    $ \\
  $\text{tSQ}_2$ Nish. & 1.714 &   $0.89081$    &  $0.9528    $ \\
 $\text{SSQ}_2$   & 2.352 &   $0.827$    &  $1.277    $ \\
 $\text{SQ}_4$  & 2.391 &  $0.835 $ &  $1.231    $ \\
 $\text{SSQ}_4$ &  2.802 &$0.809$ &    $1.388    $  \\
 2D \cite{Onsager43, Pelissetto08, Ohzeki09c} ~~  & ~~$2.269...$~~   & ~~$0.89081(7)$~~ & ~~$0.9528(4)$ \\
\hline
\hline
\end{tabular}
\caption{ Estimate of the FM critical point ($p=1$) temperature
  ($T_\text{Ons}$) and the coordinate of intersection point between
  the PM/FM transition line with the Nishimori line ($p_{\text{mc}}$,
  $T_{\text{mc}}$) for the disordered bimodal 2D Ising model obtained
  with the different block RG transformations discussed in this work,
  compared with the locations known for the 2D lattice.  }
\label{tab:mc_point2D}
\end{center}
\end{table}

\subsection{$4$-{\em staggered} cells cluster (``$\text{SSQ}_4$'')
}

As found for the $\text{SQ}_2$ cluster, the $\text{SQ}_4$ cluster does
not show an AFM fixed point and the PM/AFM transition is missing.  To
recover it we then consider the generalization to a \emph{staggered}
grouping of spins for the four cells cluster (``$\text{SSQ}_4$'').  By
referring to the numbering of Fig. \ref{4celle_2d} :
\begin{align*}
 \{ s_1, s_5, s_9, s_{13} \} &\rightarrow s_a \, ,
 \\
 \{ s_2, s_6, s_{10}, s_{14} \} &\rightarrow s_b \, ,
 \\
 \{ s_3, s_7, s_{11}, s_{15} \} &\rightarrow s_c \, , 
 \\
 \{ s_{4}, s_{8}, s_{12}, s_{16} \} &\rightarrow s_d  \, .
\end{align*}
The  phase diagram obtained with this block RG transformation is shown in Fig. 
\ref{fig:Ising2D_Tp_PhDi}.
Though we can now identify the PM/AFM
transition, we observe a worsening of the estimates of the
critical points: $T_{\text{c}}=2.802$ for both the Curie and the N{\'e}el points.
The points along the PM/FM transition line flow towards the FM fixed
point at $p=1$, while those on the PM/AFM transition line are
attracted by the AFM fixed point at $p=0$.  Therefore also in this
case we do not find a strong disorder fixed point.

The two relevant scaling exponents of the stability matrices at the FM
critical fixed point are $y_T =0.8177$ and $y_h=1.609$; see Table
\ref{tab:2d_2celle_BW} for the corresponding critical exponents.

For the AFM fixed point we have $y_T =0.8177$, the same of the FM
fixed point.  As discussed previously, for the AFM transition the
relevant order parameter is the {\em staggered} magnetization, and the
scaling exponent of the {\em staggered} field is $y_{h^\dagger} =
1.569$.

To summarize, for the 2D Ising model with bimodal disorder, 
Eq. (\ref{f:P_J}), we have evidence for both PM/FM and 
PM/AFM transition for large enough $|p|$.  Quantitatively, the
best estimates for the Curie and N{\'e}el critical points are obtained
in the $\text{SSQ}_2$  cluster scheme (cfr. Table \ref{tab:mc_point2D}). 
The multi critical point is missing since we do not find any strong disorder fixed point.
No SG phase can be tested because we are in dimension $d<2.5$. 
Therefore in the next Section we move to the 3D case.


\section{Cluster Renormalization Group for the 3D Ising model}
\label{s:Ising3D}

\begin{figure}[t!]
\centering
\includegraphics[width=\columnwidth]{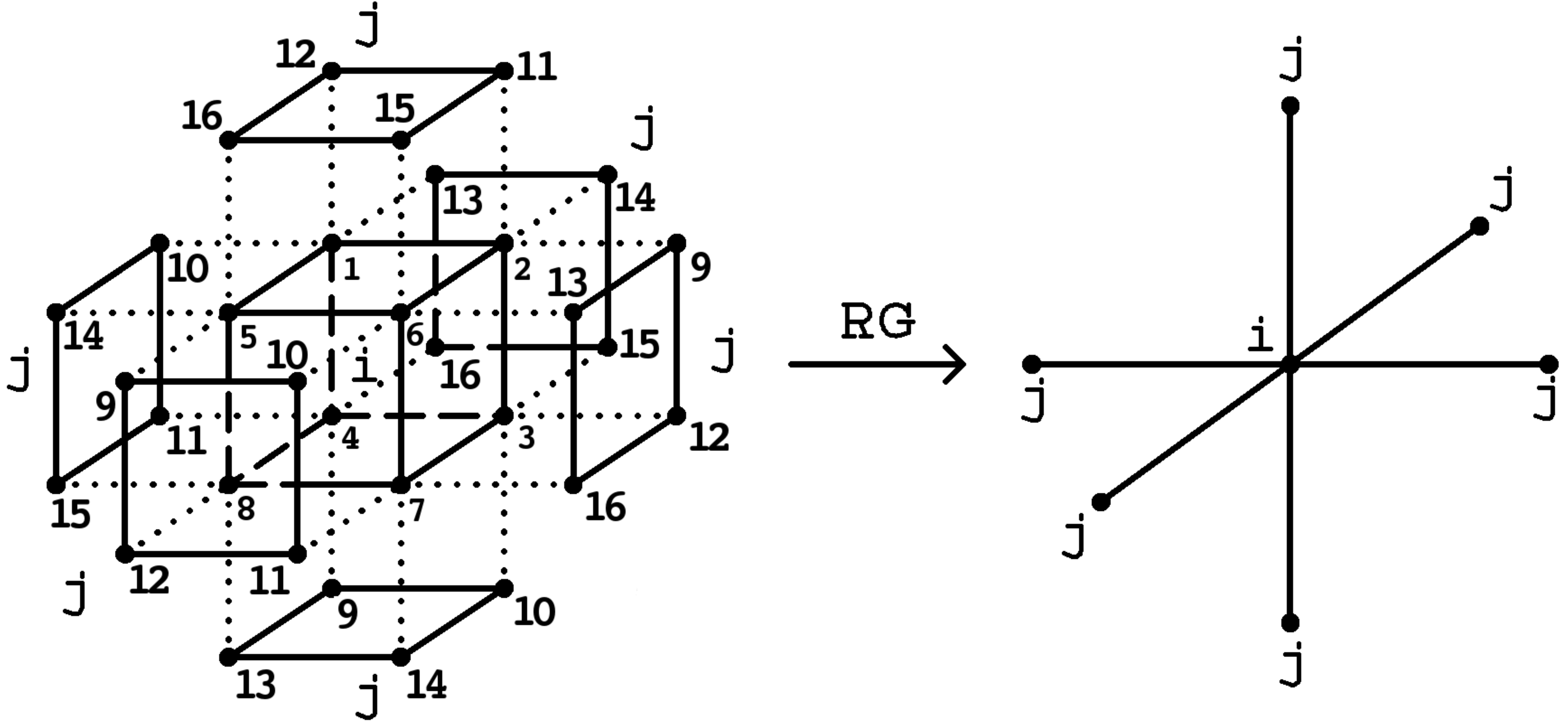}
\caption{Three dimensional two cells cluster.
With the  cell grouping in figure (solid lines) we refer to it as ``CB$_2$''.}
\label{fig:cella3d}
\end{figure}

In this Section we extend the method based on the $\text{SQ}_2$ cluster to the three dimensional case,
by using  the cluster of two cubic cells with periodic boundary conditions
shown in Fig. \ref{fig:cella3d} (referred as ``$\text{CB}_2$'')
for the study of the 3D Ising model.

The associated projection matrix is 
\begin{equation*}
\begin{tabular}{cc} 
$\quad \mathcal{M}(1,s_{i\in c})  \quad$ & ${s_{i\in c}}$  \\
\hline
$1$ &  $+ + + + + + + + $ \\
$1-t_6$ &  $+ + + + + + + - $ \\
$1-t_4$ &  $+ + + + + + - - $ \\
$1-t_2$ &  $+ + + + + - - - $ \\
$1/2$ &  $+ + + + - - - - $ \\
$t_6$ &  $+ + + - - - - - $ \\
$t_4$ &  $+ + - - - - - - $ \\
$t_2$ &  $+ - - - - - - - $ \\
$0$ & $--------$
\end{tabular}
\end{equation*}
and  $\mathcal{M}(-1,-s_{i\in c})=\mathcal{M}(1,s_{i\in c})$, which, for $t_i=0$, reduces to the majority rule.

\begin{figure}[t!]
\centering
\includegraphics[width=\columnwidth]{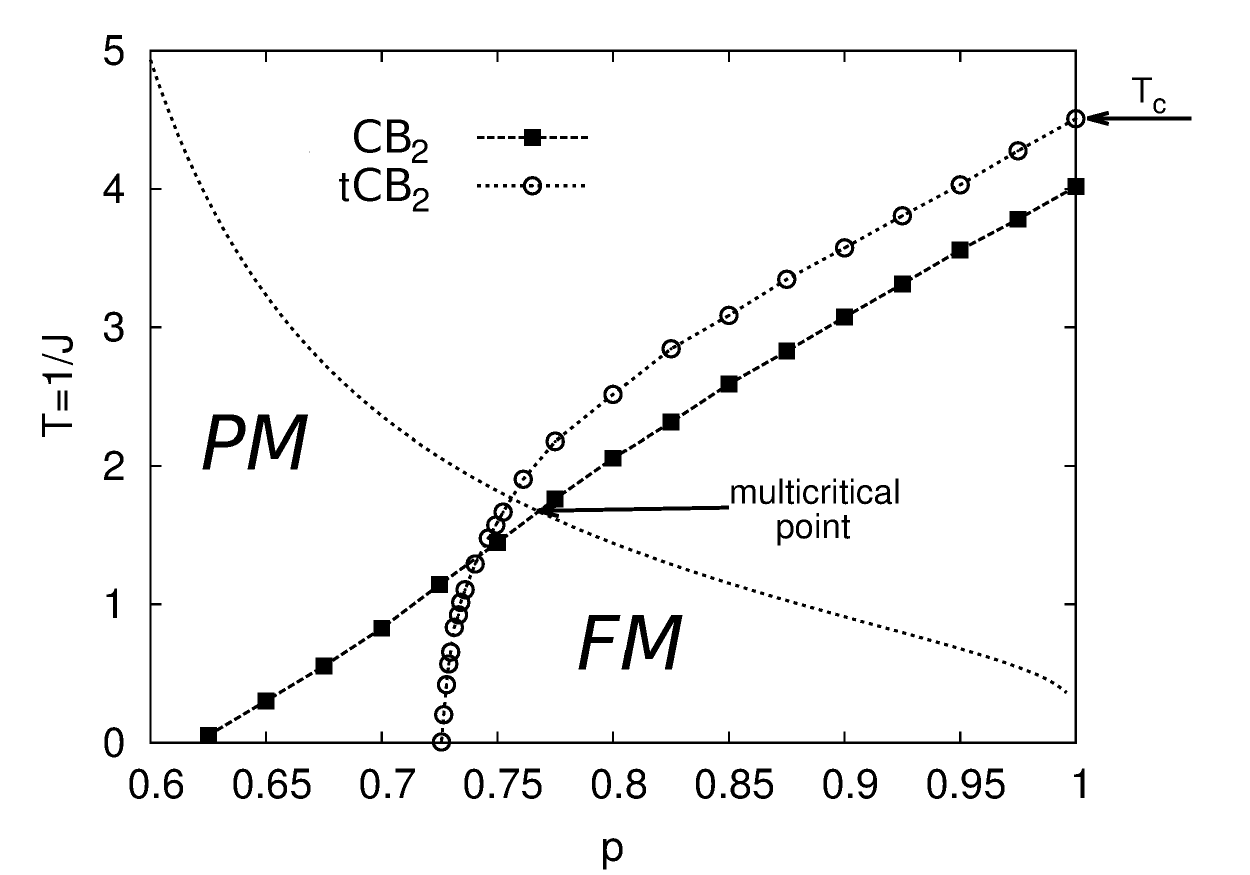}
\caption{Phase diagram in the $(p,T)$ plane of the $\pm J$ 3D Ising model 
                obtained with the block  RG transformation
                using the two cells clusters CB$_2$ and tCB$_2$ (see text).
                The dashed line is the Nishimori line.
                The $t\not=0$ curve is obtained by fixing the values $t_i$ by 
                 the requirement  $T_{\text{c}}(p=1) = 4.5115$.
   }
\label{fig:PhDi_3D_BWvneq0}
\end{figure}

The initial probability distribution of the interactions is given in
Eq. (\ref{f:P_J}), and we used $N_s=10$ pools of size $M=10^6$.  The
phase diagram for $\text{CB}_2$ cluster is shown in
Fig. \ref{fig:PhDi_3D_BWvneq0}.  Once again, only the pure fixed point
at $p=1$ controlling the PM/FM transition is found.

For the choice $t_i=0$ the critical temperature is $T_{\text{c}}=
4.0177$, which compared with the estimation from numerical simulations
$T_{\text{c}}=4.5115$ \cite{Talapov96}, has a difference of about
$12\%$.  The scaling exponents of the fixed point are $y_T =1.253$ and
$y_h=2.684$, the value of the critical exponents are reported in Table
\ref{tab:3d_ce}.
\begin{table}[t!]
\begin{center}
\begin{tabular}{lcccccc}
\hline
\hline
           & $\boldsymbol \alpha$  & $\boldsymbol \beta$ & $\boldsymbol \gamma$ & $\boldsymbol \delta$ & $\boldsymbol \nu$ & $\boldsymbol \eta$  \\
\hline
CB$_2$               & -0.3952   & 0.2521  & 1.891    & 8.499    & 0.7984 & -0.3684  \\
tCB$_2$ & -0.3015   & 0.4413  & 1.419    & 4.215    & 0.7672 & 0.1505  \\
SCB$_2$ & -0.8887   & 0.4944  & 1.900    & 4.843    & 0.9629 & 0.02693  \\
3D \cite{Pelissetto02} ~~ &  0.1101   &  0.3265  & 1.2373   & 4.789  & 0.6301   &  0.03645   \\
\hline
\hline
\end{tabular}
\caption{FM critical exponents of the 3D Ising model obtained with the block  RG transformation
                using the two cells clusters discussed in the text.
                For the tCB$_2$ method the values of $t_i$ are fixed by the requirement 
                $T_{\text{c}}(p=1) = 4.5115$.
  }
\label{tab:3d_ce}
\end{center}
\end{table}
The PM/FM transition line crosses the Nishimori line at the point
$p_{\text{mc}}= 0.76793$ and $T_{\text{mc}}=1.6721$, compatible with
the multicritical point obtained for the 3D Ising model on a the cubic
lattice \cite{Ozeki98}: $p_{\text{mc}}= 0.7673(4)$,
$T_{\text{mc}}=1.676(3)$.  Despite this agreement the transition line,
however, does not show any re-entrance.

When the parameters $t_i$ are fixed by the condition
$T_{\text{c}}(p=1)=4.5115$ \cite{Talapov96}, leading to $t_2 = 0.011$,
$t_4=-0.010$ and $t_6 = -0.050$, the transition line shows a sharp
increase of the slope for $p<p_{\text{mc}}$, yet no re-entrance nor
vertical part are observed, see tCB$_2$ line in
Fig. \ref{fig:PhDi_3D_BWvneq0}.  There is still only the critical
point at $p=1$ and the crossing with the Nishimori line does not
correspond to a real multicritical point.  The scaling exponent are
$y_T=1.303$ and $y_h = 2.425$, and one observes a slight improvement
of the values of the critical exponents, see Table \ref{tab:3d_ce}.

The condition $T_{\text{c}}(p_{\text{mc}}) = T_{\text{mc}}$ is
compatible with $t_i=0$ and does not give new results.

The 3D Ising model is known to present a multicritical point where PM,
FM and SG phase meet \cite{Nishimori81}.
Contrarily to numerical simulation predictions, where
$\alpha=2-d\nu>0$ \cite{Pelissetto02}, in both cases we find a
negative $\alpha$, indicating that in the RG analysis the FM $p=1$
fixed point is stable against the introduction of quenched disorder.
Indeed, as noted above, the RG based on the $\text{CB}_2$ cluster
fails to locate any fixed point different from the PM/FM $p=1$
critical fixed point.

\subsection{$2$-{\em staggered} cubic cells cluster (``$\text{SCB}_2$'')}
\begin{figure}[b!]
\centering
\includegraphics[width=\columnwidth]{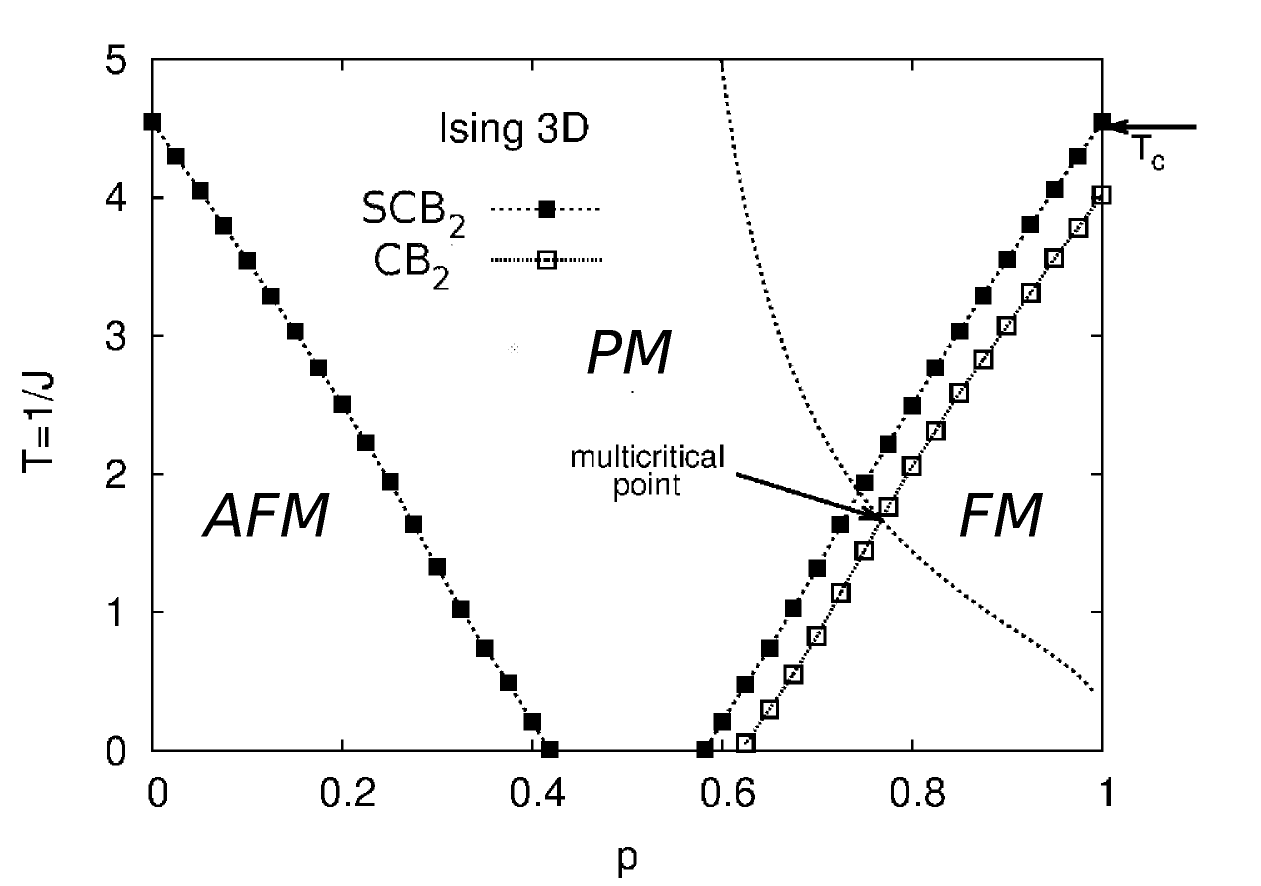}
\caption{Phase diagram in the $(p,T)$ plane of  the $\pm J$ 3D Ising model
                obtained using the $2$-cubic cell cluster with $t_i=0$, line CB$_2$, and
                the $2$-staggered cell cluster, line SCB$_2$.
                 The dashed line is the Nishimori line.
               }
\label{fig:Ising3D_Tp_PhDi}
\end{figure}

As done for the 2D case, to locate the AFM fixed point we modify the
two cells cluster of Fig. \ref{fig:cella3d} to have a staggered
topology, and hence preserving a possible antiferromagnetic ordering
in the renormalization process.
The resulting phase diagram 
is shown  in Fig. \ref{fig:Ising3D_Tp_PhDi}, line ``$\text{SCB}_2$''. 
Now, besides the FM fixed point at $p=1$, a symmetric AFM fixed point
at $p=0$ appears. The critical temperature is $T_{\text{c}}=4.5537$, closer
to the FM critical temperature found from from numerical simulations 
(cfr. Table \ref{tab:mc_point3D}).
The scaling exponents for the FM fixed point are $y_T =1.039$ and
$y_h=2.487$, see Table \ref{tab:3d_ce} for comparison of the
corresponding critical exponents.  The exponent $\alpha$ is negative,
even more than the previous cases, signaling the absence of other
fixed points for $1/2\leq p< 1$, according to the Harris criterion
\cite{Harris74}.

For the AFM critical fixed point we get $y_T =1.039$ and $
y_h=0.6075$, while that of the \emph{staggered} magnetic field is
$y_{h^\dagger} = 1.487$.

The re-entrance of the transition line below the Nishimori line is
missing, confirming also for the 3D case the limitations of the
block RG transformation based on the small cluster  scheme.

In conclusion, the phase diagram obtained for the 2D and 3D Ising
models are qualitatively similar, with the notable absence of any SG
phase in the 3D case.

The extension to larger cells, similar to the one discussed in
Sec. \ref{ss:4c} for the 2D case, becomes readily unfeasible for 3D
lattices.  For example with $8$ cubic cells one should sum over the
configurations of $4^3$ spins, more than $10^{14}$ times the
configurations of the two cells cluster.

However, based on the results of the 2D case, we do not expect that
such an extension would solve the problem of the SG phase.  To catch
the SG phase one has to look for different block RG transformation
strategies that account for SG local order.  In particular, spins
could not be right variables to be directly mapped in the RG
procedure, since the local magnetization is not a meaningful parameter
for the SG phase.

\begin{table}[t!]
\begin{center}
\begin{tabular}{lccc}
\hline
\hline
   & ~ \bfseries T$_{\text{c}} \mathbf{(p=1)}$ ~  & \bfseries p$_{\text{mc}}$ & \bfseries T$_{\text{mc}}$ \\
\hline
CB$_2$      &   4.0177           &\,  $0.7679\phantom{(x)}$  & $1.672\phantom{(x)}$  \\
tCB$_2$     &  4.5115   &\,  $ 0.7562\phantom{(x)}$ & $ 1.767\phantom{(x)}$ \\
SCB$_2$    &  4.5537 &\,  $0.7445\phantom{(x)}$ & $1.870\phantom{(x)}$    \\
3D \cite{Ozeki98}\, ~ & 4.5115  & \, $0.7673(4)$     ~~    & $1.676(3)$ \\
\hline
\hline
\end{tabular}
\caption{
Estimate of the FM critical fixed point and of the 
intersection between the PM/FM transition line with  
                the Nishimori line for the disordered bimodal 3D Ising model
                obtained using the two cell clusters discussed in the text.  
                In the last line we compare with the values for the 3D Bravais lattice.
} 
 \label{tab:mc_point3D}
\end{center}
\end{table}

\section{Cluster Renormalization Group for the Blume-Emery-Griffiths model}
\label{s:BEG}

In this Section we apply the RG analysis to the BEG model, a spin-$1$
model introduced for the study of the superfluid transition in
He$^3$-He$^4$ mixtures \cite{Blume71}.
The BEG model was originally studied in the mean-field approximation
in Refs. [\onlinecite{Blume66,Capel66,Blume71}].  Finite dimensional
analysis has been carried out by different means, e.g., series
extrapolation techniques \cite{Saul74}, RG analysis \cite{Berker76},
Monte Carlo simulations \cite{Deserno97}, effective-field theory
\cite{Chakraborty84} or two-particle cluster approximation
\cite{Baran02}.
Extensions to quenched disorder, both perturbing the ordered fixed
point and in the regime of strong disorder, have been studied
throughout the years by means of mean-field approximation
\cite{Crisanti02,Crisanti04, Crisanti05}, real space RG analysis on
Migdal-Kadanoff hierarchical lattices \cite{Falicov96, Ozcelik08} and
Monte Carlo numerical simulations
\cite{Puha00,Paoluzzi10,Paoluzzi11,Leuzzi11}.

 Besides a second order phase transition, the model is known to
 display a first order phase transition associated with phase
 separation between the PM and FM phases in the ordered case, and
 between the PM and SG phases in the quenched disordered case.  This
 rich phase diagram allows for a structured analysis of the RG
 approximations.  In particular, we go through a detailed study
 of the ordered 2D BEG model, to compare with the results of
 Ref. \cite{Berker76}, and we show the main properties of the
 quenched disordered 3D BEG model, which is an relevant test model
 for RG methods of quenched disordered systems.

\begin{table}[t!]
\begin{center}
\begin{tabular}{ccc}
\hline
\hline
\bfseries Fixed point    & \bfseries  Type  &   $\mathbf{J, K, \boldsymbol \Delta}$   \\
\hline
\multicolumn{3}{c}{\it Higher-order fixed points} \\
$ \text{C}^* $                 & Critical                   & $0.4259,-0.2910,-\infty$   \\
$ \text{G}^* $                 & Critical                   &   $0, 1.701, 4.096$   \\
$ \text{L}^* $                   & Critical end           &    $0.4250, +\infty, +\infty$    \\
$ \text{T}^* $                 & Ordinary tricritical &  $0.8848, 0.9031,  3.528$ \\
$ \text{P}^* $               & Special tricritical    &  $0.4994, 1.495,  3.992$ \\
\vspace{-0.2cm} \\
\multicolumn{3}{c}{\it First-order fixed points} \\
$ \text{Fe}^* $                & Discontinuous $\mathcalligra{m}$ &  $+\infty,-\infty,-\infty$  \\
$ \text{F}^* _J, \text{F}^* _K, \text{A}^* $ & Discontinuous $\mathcalligra{m}, \, \mathcalligra{q}$ &  
                                                                       $+\infty,+\infty,+\infty$  \\
$ \text{F}^* _2 $              & Discontinuous $\mathcalligra{q}$ & $  0, +\infty, +\infty$  \\
\vspace{-0.2cm} \\
\multicolumn{3}{c}{\it Trivial fixed points} \\
$ \text{Pa}^* _+ $             & 
\begin{tabular}{@{}c@{}} Sink for $\mathcalligra{m} =0$ , \\ large $\mathcalligra{q}$ phase \end{tabular} &  
     $0,0,-\infty$  \\
$ \text{Pa}^* _- $             & 
\begin{tabular}{@{}c@{}} Sink for $\mathcalligra{m} =0$ , \\ small $\mathcalligra{q}$ phase \end{tabular} &  
$0,0,+\infty$  \\
$ \text{S}^* $                 & 
\begin{tabular}{@{}c@{}} Smooth continuation  \\ between $ \text{Pa}^* _+ $ and 
      $\text{Pa}^* _- $ \end{tabular}  &  
      $  0,0,\ln2$  \\
\hline
\hline
\end{tabular}
\caption{Location of all the fixed points of the RG flow for the 2D
  BEG model obtained with the $\text{SSQ}_2$ cluster.  The phase
  transitions are characterize by the magnetization $\mathcalligra{m}
  \equiv \langle s_i \rangle$ and the quadrupole order parameter
  $\mathcalligra{q} = \langle s_i^2 \rangle$.  The notation for the
  fixed points is the same as in Ref. \cite{Berker76}, where their
  complete description is presented.  }
\label{tab_fp_BEG_AF}
\end{center}
\end{table}

\subsection{Ordered 2D BEG model}
Following Berker and Wortis \cite{Berker76}, in the ordered case we write
the BEG Hamiltonian as
\begin{align}
- \beta \H( \{s\}) = & 
 J \sum_{\langle ij\rangle}s_is_j  + K\sum_{\langle ij \rangle}s_i^2s_j^2 
- \Delta\sum_is_i^2   \nonumber
\\
& + h\sum_is_i  + L\sum_{\langle ij\rangle} \left(s_is_j^2+s_i^2s_j\right), 
\label{eq_BEG_H}
\end{align}
where $s_i = 0, \pm 1$.
As for the Ising model, we are interested in the case $h=L=0$, but these interactions
must be still considered for the evaluation of the critical exponents.
All the transitions in this model are characterized by two order parameters:
the magnetization $ \mathcalligra{m} \equiv \langle s_i \rangle$ and the quadrupole 
order parameter $ \mathcalligra{q}  \equiv \langle s_i^2 \rangle$, giving the density of magnetic or 
occupied sites.

We stress that the BEG model with $h=L=0$ reduces to the Ising model discussed in the previous Section 
in two separate regions of the phase diagram:
in the limit $\Delta \to -\infty$, where the holes $s_i=0$ are trivially suppressed;
and on the $J=0$ plane, where the magnetization is zero and the model reduces to the Ising model
for the spin variable $ u_i \equiv 2 s_i^2-1$  in the field $h_u = K+(\ln 2 -\Delta)/2$.
The exact mapping between the region $\Delta \ll -1 $ and the line $J=0, \, \Delta = 2 K + \ln 2$
is known as \textit{Griffiths symmetry} \cite{Griffiths}. A basic requirement for our RG transformation
is, consequently, to be equivalent in the two regions and to reduce to the one previously defined for the Ising model.

We shall consider the block RG transformation based on the same clusters used for the 2D  Ising model, 
and in particular for the $\text{SQ}_2$ cluster we reproduce results coinciding with those of Ref. \cite{Berker76}.

The generalization of the cell projection matrix of Sec. \ref{sec:2DIFM} to the spin-1 case is provided by
\begin{equation*}
\begin{tabular}{cccc}
$ \mathcal{M}(1,s_{i\in c})$\, & $\mathcal{M}(-1,s_{i\in c})$\, & $\mathcal{M}(0,s_{i\in c})$\, & ${s_{i\in c}}$  \\
\hline
$1$                        &  $0$                         & $0$                       &  $+ + + + $ \\
$1-t$                     & $t$                          & $0$                       &  $ + + + - $ \\
$1/2$                     & $1/2$                      & $0$                       &  $ + + - - $ \\
$t$                         &$1-t$                       & $0$                       &  $ + - - - $ \\
$0$                        &  $1$                         & $0$                       &  $- - - - $ \\
\vspace{-0.2cm} \\
$1-t$                    &   $0$                        &$t$                        &   $+ + + \, 0$ \\
$1-t$                    &   $0$                        & $t$                       &   $ + + - \, 0$ \\
$0$                       & $1-t$                       & $t$                       &   $ + - - \, 0$ \\
$0$                       & $1-t$                       & $t$                       &   $ - - - \, 0$ \\
\vspace{-0.2cm} \\
$1/2$                   & $0$                           &$1/2$                   &   $+ + 0 \, 0$ \\
$1/4$                   & $1/4$                       & $1/2$                   &   $+ - 0 \, 0$ \\
$0$                      & $1/2$                       & $1/2$                   &   $- - 0 \, 0$ \\
\vspace{-0.2cm} \\
$t $                     &$0$                           & $1-t$                   &    $+ \, 0 \, 0\,  0$ \\
$0$                      &$t$                           & $1-t$                   &    $-  \, 0 \, 0 \, 0$ \\
\vspace{-0.2cm} \\
$0$                     & $0$                          & $1$                       &   $ 0 \, 0 \, 0 \, 0 $ \\
\end{tabular}
\end{equation*}
This is the most general cell projection matrix that contains the up-down, the Griffiths and the square symmetries \cite{Berker76}.
In particular, for $t=0$ it reduces to the {\em double majority rule}: the majority rule is first applied to the variable $u_i \equiv 2 s_i^2 -1$,
and then, if the magnetic sites are dominant, to $s_i = \pm 1$.

The block RG transformation leads to the renormalized Hamiltonian for the new spin variables
\begin{align}
-\beta {\cal  H}_R(s_a,s_b) = \alpha \, \Bigl[& J_R\, s_a s_b + K_R\, s^2 _a s^2 _b + 
\\
 & + L_R (s_a ^2 s_b +s_a s_b ^2) + \nonumber
 \\
&- \Delta_R (s_a^2 +s_b^2) + h_R (s_a + s_b)\Bigr] \nonumber
\end{align}
with
\begin{align}
 J_R =& \frac{1}{4\alpha} \log \biggl( \frac{x_{++} \, x_{--}}{x_{+-}^2} \biggl)   \nonumber
\\
 K_R =& \frac{1}{4\alpha} \log \biggl( \frac{x_{++}\, x_{--}\, x_{+-}^2\, x_{00}^4}{x_{+0}^4\,  x_{-0}^4} 
                                                      \biggl)  \nonumber
\\
 \Delta_R =& \frac{1}{2} \log \biggl( \frac{x_{00}^2}{x_{+0}\, x_{-0}} \biggl) 
\\
 L_R =& \frac{1}{4\alpha} \log \biggl( \frac{x_{++}\, x_{-0}^2}{x_{--}\, x_{+0}^2} \biggl)   \nonumber
\\
 h_R =& \frac{1}{2} \log \biggl( \frac{x_{+0}}{x_{-0}} \biggl)  \nonumber
\end{align}
where $x_{s_a s_b}$ are the edge factors (\ref{f:Boltzmann}) and $\alpha=4$ for the 2D lattice.
Note as in our case $h=L=0$ at the beginning and they are not generated in the RG process.
The explicit expressions for $h_R$ and $L_R$ are nevertheless required to obtain the critical exponents (cfr. Appendix \ref{app:SM_BEG}).

The evaluation of the stability matrix can be problematic 
if the RG flux flows towards a fixed point where one of the
parameters is infinite, e.g., $\Delta\to-\infty$. 
In cases like this it is more convenient to use a variable 
remaining finite at the fixed point, e.g., $A=e^\Delta$.

The locations of all the fixed points in the RG flow generated by the block RG based on 
the $\text{SSQ}_2$ cluster are reported in Table \ref{tab_fp_BEG_AF}.

The fixed points  $\text{C}^*$, $\text{G}^*$ and $\text{P}^*$ are of particular interest for testing 
 the RG procedure because they are known exactly.
Moreover, the FM Ising  fixed point C$^*$  and the Griffiths fixed point G$^*$ are related to each other.
The first occurs for $\Delta\to-\infty$, while the second at $J=0$,
and the Griffiths symmetry \cite{Griffiths, Berker76} imposes the relations
\begin{align}
 K_{\text{G}^*} =& \,4 J_{\text{C}^*} \, , & 
 \Delta _{\text{G}^*} =& \, 8 J_{\text{C}^*} + \ln 2.
\end{align}
These relations are verified by our numerical results. 
The FM fixed point C$^*$ can be  used 
to fine tune the value of the parameter $t$  in the projection matrix, obtaining $t = -0.06453$, 
the same value found for the 2D Ising model, see Sec.\ref{s:psrgI2D}.

The completely unstable Potts fixed point P$^*$ can be used as an indicator
of the precision of the cluster approximation used in the RG analysis.
The position of the point is known to lie on the axis \cite{Berker76}
\begin{align}
 K \, =& \, 3J \, , & \Delta \, =&  \, 8J ;
\label{eq_Potts_Axis}
\end{align}
where the Hamiltonian (\ref{eq_BEG_H}) has  a three-state permutation symmetry.
On this axis the BEG model  can be reduced to the three-state Potts model
\begin{align*}
 - \beta \H = \mathcal{D} \sum_{\langle ij \rangle}
 \left( \delta_{s_i  s_j} -1 \right) 
 \text{ with }
 \mathcal{D} = \sqrt{2 \frac{J^2 + K^2 + \Delta^2}{37}  } 
\end{align*}
and the critical point of the three-state Potts model is at $\mathcal{D} = \ln(1+\sqrt{3}) = 1.0050525...$ \cite{Berker76}.

The location of the fixed points C$^*$, G$^*$ and P$^*$ are shown in Table \ref{tab_vifp},
and we note as the $\text{SSQ}_2$ cluster gives better estimation  compared to the $\text{SQ}_2$ clusters.

In particular, the location $(J,K,\Delta)$ of the fixed point P$^*$ deviates from the exact result 
of about $(16\%,\, 16\%,\, 16\%)$ for the $\text{SQ}_2$ cluster, 
of  $(6\%,\, 2\%,\, 0.2\%)$ for the $\text{tSG}_2$ with $t = -0.06453$, 
and of about $(0.6\%, 0.8\%, 0.7\%)$ for the $\text{SSQ}_2$.
In terms of $\mathcal{D}$ this translates into 
$\mathcal{D} = 1.1696$ for the $\text{SQ}_2$ cluster, 
$\mathcal{D} = 1.001535$ for the $\text{tSQ}_2$,
 and 
$\mathcal{D} = 0.997894$ for the $\text{SSQ}_2$. 

\begin{table}[t!]
\begin{center}
\makebox[\linewidth][c]{
\begin{tabular}{cccccc}
\hline
\hline
 &   & ~~~~ \bfseries $ \text{SQ}_2$ \cite{Berker76}    & ~ \bfseries $\text{tSQ}_2$  & \bfseries ~  SSQ$_2$ ~  & \bfseries ~ 2D ~  
  \\ 
\hline
             & $J$            & 0.5275    & 0.4407 & 0.4259    & 0.4407 \\
C$^*$  & $K$          & -0.1618   & -0.2414     & -0.2910   & not known\\
              & $\Delta$ & $-\infty$ & $-\infty$  & $-\infty$   & $-\infty$
             \vspace{0.3cm} \\
             & $J$            & 0           & 0         & 0            & 0 \\
G$^*$  & $K$           & 2.110 & 1.763     & 1.701  & 1.763\\
              & $\Delta$   & 4.913 & 4.219   & 4.096   & 4.219
    \vspace{0.3cm}          \\
             & $J$         & 0.5822 & 0.5319    & 0.4994   & 0.5026 \\
P$^*$  & $K$            & 1.756   & 1.476   & 1.495   & 1.508\\
              & $\Delta$     & 4.678   & 4.012  & 3.992  & 4.020\\
\hline
\hline
\end{tabular}
}
\caption{Location of the fixed points C$^*$, G$^*$ and P$^*$ for the ordered 2D BEG model obtained with
                all the 2 cells cluster discussed in the text compared to the exact results for the 2D lattice.
                }
\label{tab_vifp}
\end{center}
\end{table}

The projection matrix defined above 
does not preserve the three-state permutation 
symmetry on the Potts-axis (\ref{eq_Potts_Axis}), as an exact RG would do.
The distance of the fixed point P$^*$ from the Potts-axis can then be used as an indicator
of the error made with the cluster approximation used to build the block RG transformation.
The distance of $\text{P}^*$ from the Potts-axis, over 
 its distance from the origin, 
turn out  to be $6 \times 10^{-4}$  for the $\text{SQ}_2$ cluster,
$10^{-2}$ for the tuned $\text{tSQ}_2$ cluster 
and $4 \times 10^{-4}$ for the $\text{SSQ}_2$ cluster.
Note, specifically, that a strong violation is obtained with the $\text{tSQ}_2$ cluster with the
``unphysical" negative $t$.

Finally, in Table \ref{tab_ce_GCLP} we show the five scaling exponents for the fixed points $\text{G}^*$, $\text{C}^*$, $\text{L}^*$ and $P^*$
(cfr. Appendix \ref{app:SM_BEG}).
We stress as the critical exponents obtained with the SSQ$_2$ cluster approximation 
are more precise than those obtained with the original square cells cluster SQ$_2$. 

Using, alternatively, the free $t$ trick, the critical exponents
are more similar to the known exact ones respect to the \emph{staggered} cells cluster case. 
Especially, the exactly known exponents for 
$\text{C}^*$, $\text{G}^*$ and $\text{L}^*$ are considerably better approached 
with the tSQ$_2$ cluster. 
This is not surprising since $t=-0.06453$ fixes
 the exact location for $\text{C}^*$ (and $\text{G}^*$),
and  we, then, expect that also the estimates of their scaling exponents improve.
The known exponents of P$^*$ show, instead, only a slight improvement.

\begin{table}[t!]
\begin{center}
\makebox[\linewidth][c]{
\begin{tabular}{lccccc}
\hline
\hline
 & \bfseries  & ~ ~ \bfseries SQ$\mathbf{_2}$ \cite{Berker76} & ~ \bfseries tSQ$\mathbf{_2}$  & ~~ \bfseries SSQ$\mathbf{_2}$ ~~  & ~~ \bfseries 2D \\
\hline
      & $y_2$   & 0.7267 & 0.9419 &  0.7534 & 1\\
      & $y_4$  & -1.0492 & -1.644 & -0.2714 & \\
$\text{C}^*$ & $y_6$ & $-\infty$ & $-\infty$ & $-\infty$ & \\
      & $y_1$  & 1.942 & 1.870 & 1.839 &1.875 \\
      & $y_3$  & 0.3792 & -0.3556& 0.3408 & 
\vspace{0.3cm} \\
      & $y_2$ &  0.7267 & 0.9419 &0.7534 & 1 \\
      & $y_4$ &  1.942 & 1.870 & 1.839 & 1.875\\
$\text{G}^*$ & $y_6$ &  -1.834 & -1.638 &-0.8473 & \\
      & $y_1$ &  0.5748 & 0.6628 & 0.5501  & \\
      & $y_3$ &  -0.7327 & -0.6731 & -0.5270 & 
\vspace{0.3cm} \\
      & $y_2$  & 0.7267 & 0.9419 & 0.7534 &  1\\
      & $y_4$ & 2.000 & 2.000 & 2.000 & \\
$\text{L}^*$ & $y_6$ & $-\infty$ & -0.5095 & $-\infty$  & \\
      & $y_1$   & 1.942  & 1.870 & 1.839 & 1.875 \\
      & $y_3$  & 0.2355 & -0.3208 & 0.3428 &
\vspace{0.3cm} \\
      & $y_2$  & 1.942 & 1.870 & 1.854 & $1.8\overline{6}$ \\
      & $y_4$  & 0.8327 & 1.106 & 0.8958 &  1.2 \\
$P^*$ & $y_6$  & 0.4645 & 0.5248 & 0.4383 & \\
      & $y_1$  & 1.936 & 1.869 & 1.837 & \\
      & $y_3$  & 0.3846 & 0.5304 & 0.3021 & \\
\hline
\hline
\end{tabular}
}
\caption{      
  Scaling exponents of the fixed points $\text{C}^*$, $\text{G}^*$, $\text{L}^*$ and
  $\text{P}^*$ obtained by means of different cell clusters.  
  The parity of the scaling exponent index refers to the parity of the interaction.
  The exponent $y_{2C} = y_{2G} = y_{2L} =1 $ corresponds to the thermal eigenvalue of the Onsager transition ($y_T$), 
  while the exponent $y_{4G} = y_{1C}=y_{1L}=1.875$ corresponds to the magnetic eigenvalue one ($y_H$) \cite{Berker76}.  
  The exact critical exponents for the $\text{P}^*$ fixed point correspond instead to the transition 
  in the three-state Potts model \cite{PottsRev}.
  }
\label{tab_ce_GCLP}
\end{center}
\end{table}

\subsection{3D BEG with quenched disorder}
In this Section we extend the analysis to the quenched disordered BEG model in
three dimensions. 
The quenched disordered 3D BEG model represents a relevant test for
the cluster RG applied to  disordered systems.
Monte Carlo numerical simulations \cite{Paoluzzi10} show a critical transition line between 
the PM phase and a SG phase, which, similar to what found in the mean-field study \cite{Crisanti02},
consists of a second order transition terminating in  
a tricritical point  from which a first order inverse transition starts. 
Furthermore, a re-entrance of the first order transition line 
is present for positive, finite values of the chemical potential of the holes \cite{Crisanti05}, 
yielding the so-called inverse freezing phenomenon.
The real space RG  study of Ozcelik and Berker \cite{Ozcelik08} based on Migdal-Kadanoff cells 
does not  reveal any first order phase transitions, nor any re-entrance.
When the real space RG is extended to more structured hierarchical lattices \cite{ACL_HL}
the  re-entrance can be recovered, 
but no tricritical point and first order transition are found.

The Hamiltonian of the disordered  BEG model suitable for the RG study is
\begin{align}
 - \beta  \mathcal{H} =  &
+ \sum_{\left\langle ij \right\rangle}  J_{ij}s_i s_j 
+ \sum_{\left\langle ij \right\rangle} K_{ij} s^2_i s^2_j + \nonumber
\\
& - \sum_{\left\langle ij \right\rangle} \Delta_{ij} \left( s^2_i + s^2_j \right) 
- \sum_{\left\langle ij \right\rangle} \Delta_{ij}^{\dagger} \left( s^2_i - s^2_j \right)
\end{align}
where the couplings are quenched random variables with the probability distribution
\begin{align}
 P(\boldsymbol{\mathcal{K}}_{ij})  =&  
 \bigl[ (1-p)\, \delta (J_{ij} + J) + p\, \delta (J_{ij} - J) \bigr]  \nonumber
 \\
 & \times
 \delta  \left( K_{ij} -K \right)  
 \delta  \left( \Delta_{ij} - \Delta \right) 
 \delta  \left( \Delta^\dagger_{ij} \right).
 \label{f:P_J_BEG}
\end{align}
If an external field $h$ is added, besides the single site term,  one has to include also the odd 
interaction term $s_i s_j^2$. 

The model has been studied using the $\text{CB}_2$ cluster shown in Fig. \ref{fig:cella3d} 
and its staggered version $\text{SCB}_2$ using in both cases $N_s=10$ pools of size $M=10^6$.

\begin{figure}[t!]
\centering
\includegraphics[width=\columnwidth]{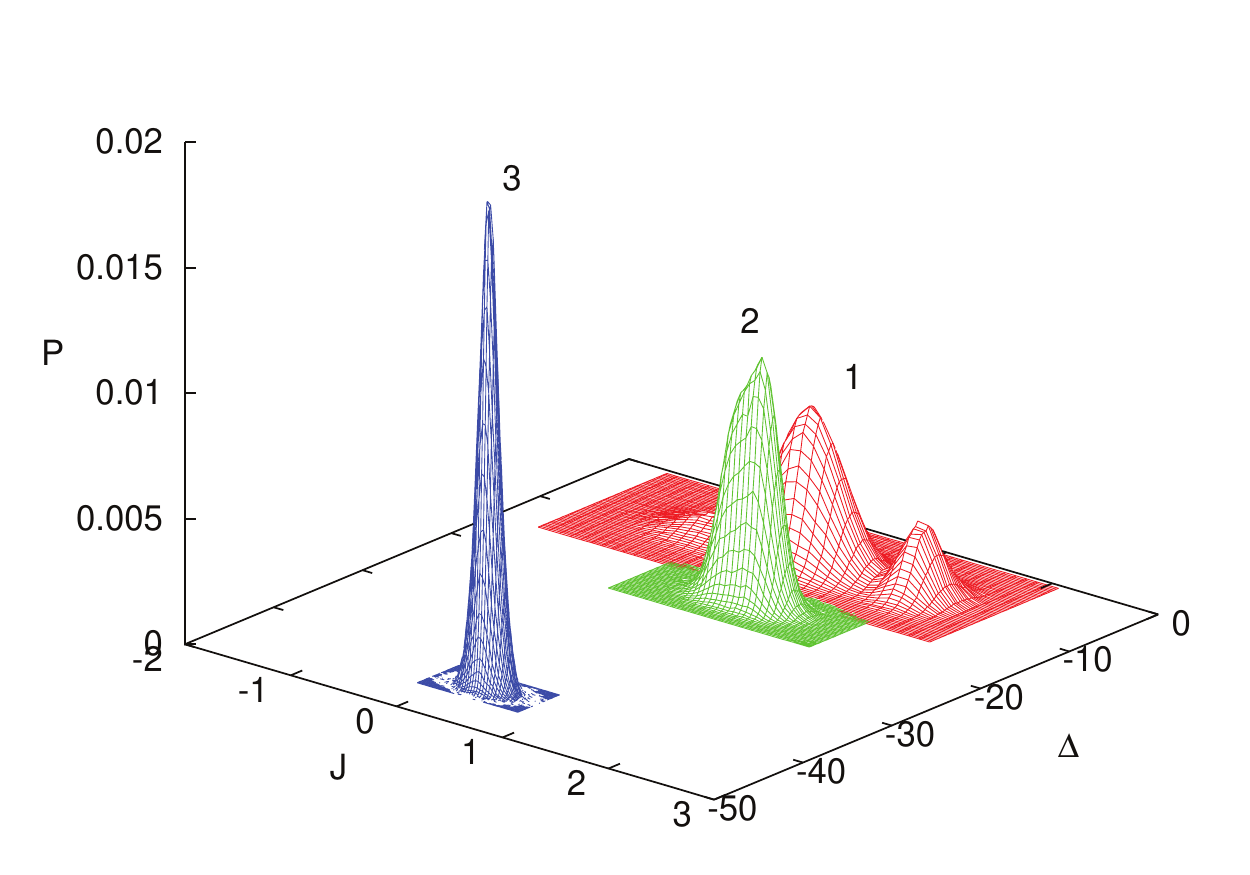}
\caption{
Flow of the renormalized probability distribution $P(\boldsymbol{\mathcal{K}})$
for the disordered 3D BEG model in the paramagnetic phase:
$J=4$, $K=0$, $\Delta=0.4$ and $p=0.6$ on the SCB$_2$ cluster. 
The parameters $K$ and $\Delta^\dagger$ are integrated.
}
\label{fig:BEG_EVO_PM}
\end{figure}

\begin{figure}[t!]
\centering
\includegraphics[width=\columnwidth]{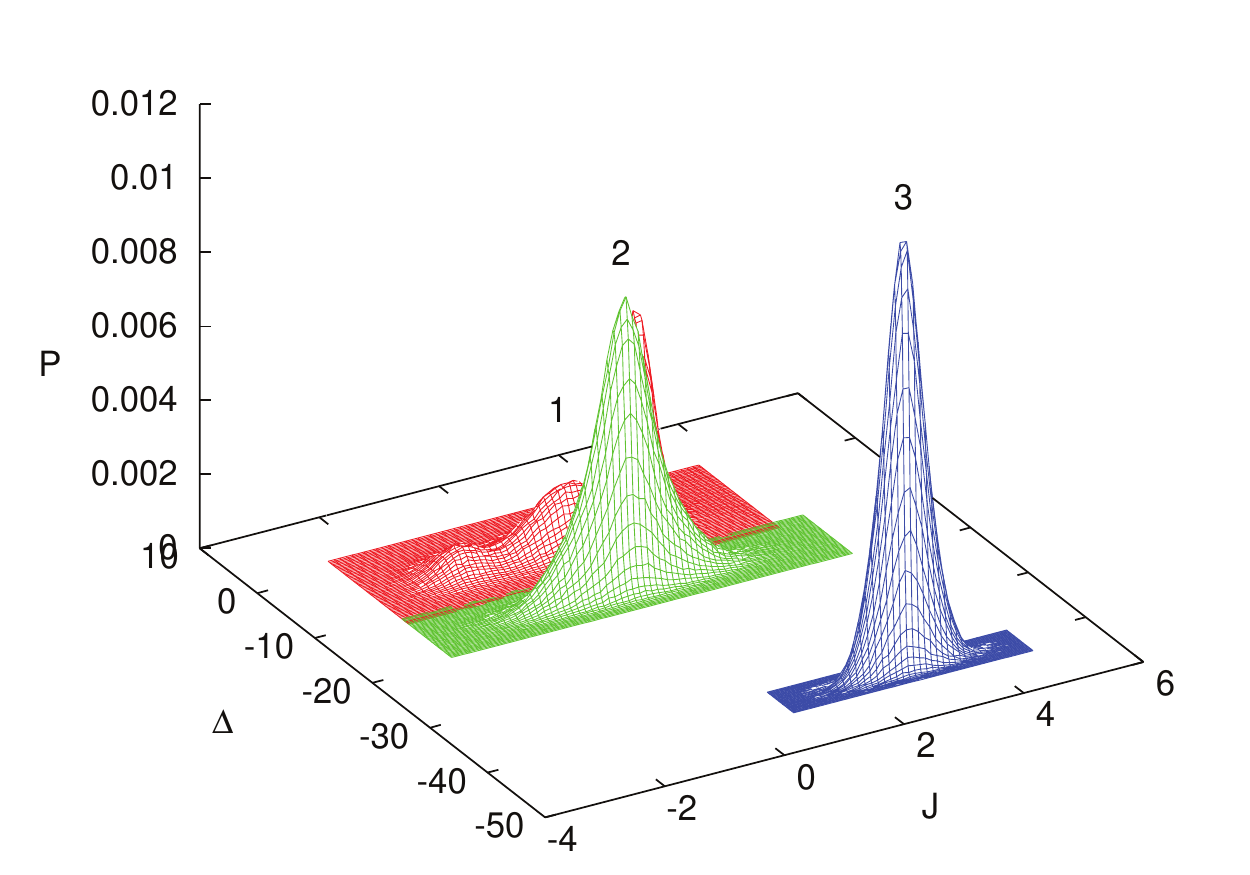}
\caption{Flow of the renormalized probability distribution $P(\boldsymbol{\mathcal{K}})$
                for the disordered 3D BEG model in the ferromagnetic phase:
                $J=4$, $K=0$, $\Delta=0.4$ and $p=0.7$ on the SCB$_2$ cluster. 
                The parameters $K$ and $\Delta^\dagger$ are integrated.
}
\label{fig:BEG_EVO_FM}
\end{figure}
Similar to what seen in the previous Section, only the PM and the FM phases are found, 
while the  SG phase remains undetected in the whole phase diagram. 
Two typical flows of the probability distribution towards the PM and FM fixed points 
are shown  in Figs. \ref{fig:BEG_EVO_PM} and \ref{fig:BEG_EVO_FM}.
In the PM phase the average value of $J_{ij}$ goes to zero, while in the FM it moves towards  $+\infty$.
In both cases the distributions become narrower and narrower under the block RG 
transformation.

\begin{figure}[t!]
\centering
\includegraphics[width=\columnwidth]{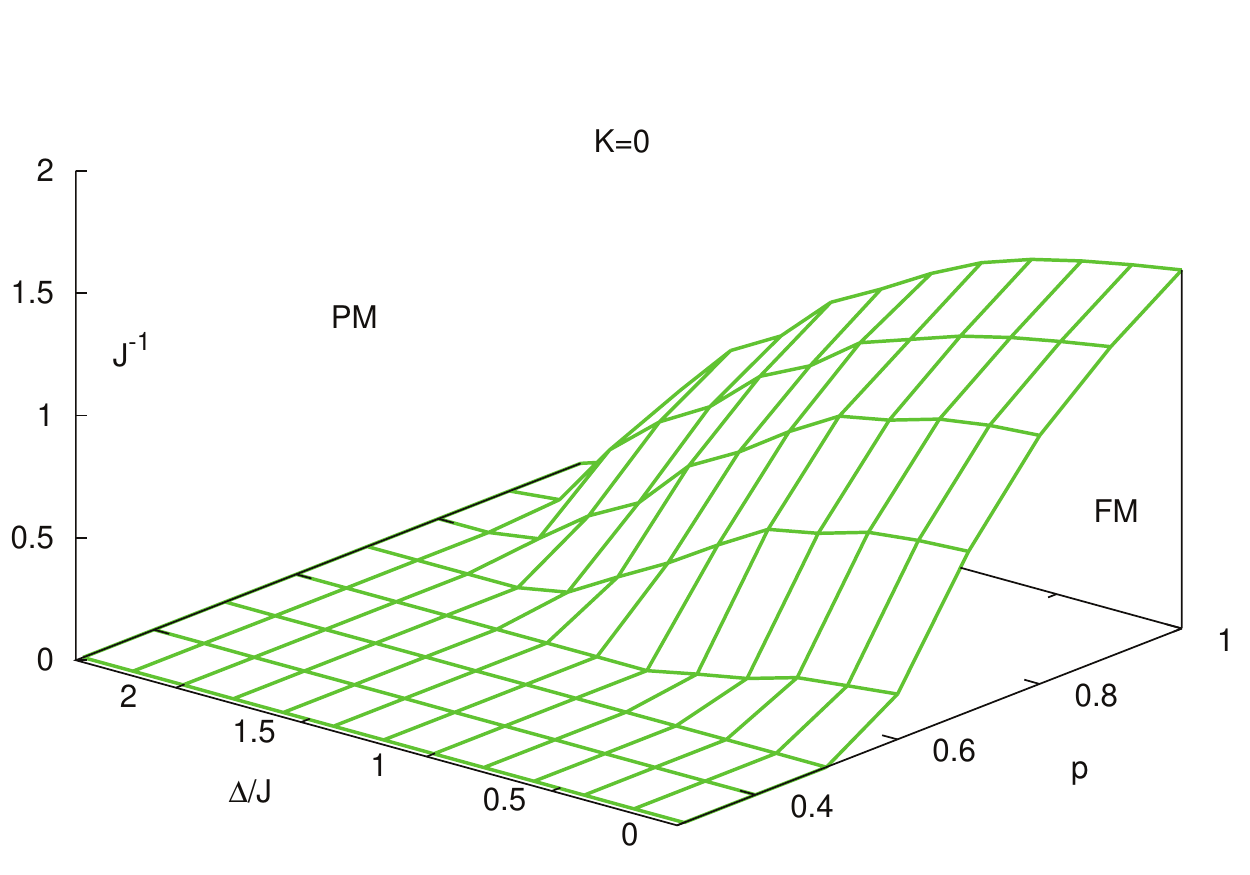}
\caption{PM/FM critical surface in the $(T,\Delta/J, p)$ parameter space  for 
                 the 3D BEG model with $K=0$
                 obtained with the $\text{SCB}_2$ cluster. 
}
\label{fig:BEG_PD}
\end{figure}

The PM/FM critical surface in the space $(T,\Delta/J,p)$ for the $K=0$ case obtained with the
$\text{SCB}_2$ cluster is shown in Fig. \ref{fig:BEG_PD}.
All the points on the critical surface flow under  RG towards one of the two ordered fixed points 
at $p=1$ with mean value $\mu_\Delta \to \pm \infty$ and variance $\sigma^2_\Delta \to 0$. 
The analysis of the critical properties is then reduced to the study of an ordered model.
In particular the fixed point at $\mu_\Delta = -\infty$ corresponds at the critical fixed point of the 3D Ising
model discussed in Sec. \ref{s:Ising3D}.

\section{Discussion and conclusions}
\label{s:discussion}

In this paper we have presented  an extension of the real space cluster RG method 
with two cells proposed by  Berker and Wortis \cite{Berker76}
by considering a \emph{staggered} topology for the clusters.
This not only makes the antiferromagnetic phase detectable, but  leads 
to an improvement of the estimates of the critical exponents and of the
location of the critical points for both the Ising and BEG models.

The two staggered cells cluster appears to be more reliable also with respect to the tuned version of the 
square cells cluster approach where one, or more, free parameters in the cell projection matrix are
fixed by the knowledge of some points in the phase diagrams.
The later tuning method is not only less predictive, requiring as input some known points, but 
it may lead to an ``unphysical" projection matrix  \cite{Berker76}. 
We have seen, indeed, that in certain cases, for example when fixing the critical temperature of the 
2D Ising model to the exact value, the resulting projection matrix assigns a negative contribution 
to some spin configurations to the partition sum. A choice  not providing any  physical insight.
The staggered cells cluster, instead, is physically motivated: 
the invariance of an antiferromagnetic ordering under RG.
It is remarkable that this request not only allows to study the critical properties of the Ne\`el transition, 
but quantitatively improves the results also for the pure ferromagnetic models. 

We observe as these results for the pure models are directly valid for a percolation problem:
defining an occupation variable as $\epsilon_{12} = (\sigma_1 \sigma_2 -1) /2 $,
the relative percolation threshold is achieved at $ p_c = 1/(1+e^{2 \beta_c})$.

We have then reported the results of the extension of the cell blocks RG transformation to quenched disordered systems.

We have established that in two dimensions, even in the staggered version, the results are not consistent with exact 
results for the corresponding regular lattice. 
In particular, the ferromagnetic phase is detected also beyond the intersection with the Nishimori line.
In this case we have also considered the extension to four cells cluster.  
Although the approximation is not systematic, with a square cell arrangements a clear 
improvement is achieved in the pure model. 
We observe that a four cells cluster is the minimal requirement to preserve possible 
plaquette frustration in presence of bond 
disorder under the RG process, which is necessary to identify a spin glass critical point (at $T=0$ in 2D).
Our investigation shows that the requirement, though necessary,  is non sufficient.
Indeed, the phase diagram of four cells cluster, besides  a minor improvement in the slope of the critical line, 
shows the same features of the two cells case.

In the three dimensional case, a similar scenario is obtained: 
in the pure case the \emph{staggered} version shows a clear improvement, 
while the quenched disordered extension is ineffective
and the expected spin glass phase remains undetected for both the Ising and BEG models.

This failure follows previous attempts of generalizing real space RG methods conceived for ordered 
systems to disordered systems.  The generalization to disordered
systems has led in the past to ambiguous	 results. On the one hand 
the cumulant expansion \cite{Kinzel78,Tatsumi78} has provided evidence for a spin glass phase  
in dimension $2$,  lower than the lower
critical dimension $2.5$ \cite{Franz94,Franz05b,Boettcher05}. 
On the other hand, however, the attempts to extend the block RG transformation 
on spin clusters did not yield any spin glass  fixed point, even in dimension $3$  \cite{Kinzel78}.

The lack of a spin glass phase in our scheme is also related to
the incorrect location of the boundary of the ferromagnetic phase in the disordered region.
We have shown, indeed, that  in the disordered Ising model the ferromagnetic phase enters   
also in the region forbidden by  Nishimori's gauge theory.
This occurs with all clusters used. 
The problem is only partly mitigate when the tuned cluster is used, cfr. Figs. 
\ref{fig:PhDi_2D_BWvneq0} 
and \ref{fig:PhDi_3D_BWvneq0},
and its uncontrolled nature does not allow for any further physical insight.
A milder, but more recognizable attenuation, is obtained with the four cells cluster,
cfr. Fig. \ref{fig:Ising2D_Tp_PhDi}.
In this regard we stress here that the correlation generated by the RG transformation 
among different types of couplings 
cannot be disregarded. 
In particular, taking the naive approximation $P(J,K,\Delta) \simeq P(J) P(K) P(\Delta)$, and so
using three \emph{independent} pools for the three kinds of interaction, the PM/FM transition line becomes straight, not different from what found with the SQ$_2$ cell.
Nevertheless, our analysis shows that parameter correlation is only one of the necessary ingredients, and
that the limitations of the block RG study of disordered models are not due to
the truncation process of the interactions,  but mostly to the nature itself of the block cells 
construction.

The connection between the problem in the ferromagnetic critical line and 
the detection of the spin glass phase is highlighted by looking at the single RG flow:
the variance $\sigma^2_J$ of the probability distributions  goes quickly to zero in all the detected phases.
This does not happen, for example, in the real space RG on hierarchical lattice \cite{ACL_HL}
where the variance $ \sigma^2_J $ of the couplings increases in the FM phase, 
even  though $ \sigma_J / \mu_J \to 0$, 
and the spin glass phase is detected as the region of the phase diagram where  $ \sigma_J / \mu_J \to \infty$.
It is 
clear that, in order to build a valuable generalization of the RG cluster method to strong disorder, the first 
step is to obtain the correct evolution of the FM phase for weak disorder.

Further issues take place when the extension to strong disorder is considered.
In particular, the improvement achieved with the \emph{staggered} cells clusters shows as, 
to correctly detect the antiferromagnetic phase, it is essential that the ground state of the system 
is invariant under the RG transformation. In the strong disorder regime 
this requirement becomes harder 
to satisfy, as the frustration causes a proliferation of nontrivial degenerate ground states.

The present analysis 
makes eventually clear  that, while the cell blocks RG method works 
well for pure, ferromagnetic  or 
antiferromagnetic, systems,  the generalization to the case of strong disorder calls for a different 
procedure for the block RG transformation.

The renormalization via the majority rule, or its tuned improvement, 
yields a local magnetization of the coarse grained cell.
This is meaningful as far as magnetization is the relevant order parameter of the transition. 
In the spin-glass transition, though, magnetization is zero and the relevant order parameter
 is the ``replica'' overlap.
The overlap allows, in particular, to take into account the ergodicity breaking caused by frustration, 
as it may translate into the replica symmetry breaking
of the appropriate overlap probability distribution.
To put forward a renormalization procedure based on the overlap coarse graining one has, thus, to resort to replicated clusters. 
More instances of the system should, then, be renormalized together via the value of the mutual overlap.
Such a generalization, and its numerically feasible implementation, is currently under investigation.

\acknowledgments
The research leading to these results has received funding
from  the People Programme (Marie Curie Actions) of the European Union's Seventh Framework Programme FP7/2007-2013/ 
under REA grant agreement n¡ 290038, NETADIS project and 
from the Italian MIUR under the Basic
Research Investigation Fund FIRB2008 program, grant
No. RBFR08M3P4, and under the PRIN2010 program, grant code 2010HXAW77-008.
AC acknowledge financial support from European Research
Council through ERC grant agreement no. 247328

\appendix*
\section{RG Stability Matrix for the BEG model}
\label{app:SM_BEG}
The critical exponents are obtained from the eigenvalue of the stability matrix 
$\partial \boldsymbol{\mathcal{K}}_R / \partial\boldsymbol{\mathcal{K}}$
evaluated at the fixed point $\boldsymbol{\mathcal{K}}^*$.
For the BEG model $\boldsymbol{\mathcal{K}} = \{J, K, \Delta, L, h\}$
and the elements of the stability matrix are
\begin{align*}
\frac{\partial J_R}{\partial \boldsymbol{\mathcal{K}} } \!  &= \! \frac{1}{4\alpha} \! \left( \frac{x_{++}'}{x_{++}}+\frac{x_{--}'}{x_{--}}-\frac{2x_{+-}'}{x_{+-}} \right)
\\
\frac{\partial K_R}{\partial \boldsymbol{\mathcal{K}}} \! &= \! \frac{1}{4\alpha} \! \left( \frac{x_{++}'}{x_{++}}+\frac{x_{--}'}{x_{--}}+\frac{2x_{+-}'}{x_{+-}}
\right.
\\
&\qquad \qquad\left. +\frac{4x_{00}'}{x_{00}}-\frac{4x_{+0}'}{x_{+0}}-\frac{4x_{-0}'}{x_{-0}} \right)
\\
\frac{\partial \Delta_R}{\partial \boldsymbol{\mathcal{K}}} \! &= \! \frac{1}{2} \! \left( \frac{x_{+0}'}{x_{+0}}+\frac{x_{-0}'}{x_{-0}}-\frac{2x_{00}'}{x_{00}} \right)
\\
\frac{\partial L_R}{\partial \boldsymbol{\mathcal{K}}} \! &= \! \frac{1}{4\alpha} \! \left( \frac{x_{++}'}{x_{++}}+\frac{2x_{-0}'}{x_{-0}}-\frac{x_{--}'}{x_{--}}-\frac{2x_{+0}'}{x_{+0}} \right)
\\
\frac{\partial h_R}{\partial \boldsymbol{\mathcal{K}}} \! &= \! \frac{1}{4} \! \left( \frac{x_{+0}'}{x_{+0}}-\frac{x_{-0}'}{x_{-0}} \right)
\end{align*}
where $x_{\sigma_a \sigma_b}'= \partial x_{\sigma_a \sigma_b} / \partial \boldsymbol{\mathcal{K}}$
and $\alpha = 2d$, with $d$ the space dimension.
The derivative of the Boltzmann factors can be expressed as
\begin{align*}
 \frac{\partial x_{\sigma_a \sigma_b}}{\partial J} &= \sum_{\bm{s}}  \mathcal{M}_a \mathcal{M}_b 
   \left[ \alpha \sum_{\langle ij \rangle}s_is_j \right] e^{-\beta \H(\bm{s})}
\\
 \frac{\partial x_{\sigma_a \sigma_b}}{\partial K} &= \sum_{\bm{s}}  \mathcal{M}_a \mathcal{M}_b 
    \left[ \alpha \sum_{\langle ij \rangle}s_i^2 s_j^2 \right] e^{-\beta \H(\bm{s})}
\\
 \frac{\partial x_{\sigma_a \sigma_b}}{\partial D} &= \sum_{\bm{s}}  \mathcal{M}_a \mathcal{M}_b 
       \left[ - \sum_{i}s_i^2 \right] e^{-\beta \H(\bm{s})}
\\
 \frac{\partial x_{\sigma_a \sigma_b}}{\partial L} &= \sum_{\bm{s}}  \mathcal{M}_a \mathcal{M}_b 
  \left[ \alpha \sum_{\langle ij \rangle}(s_i^2s_j+s_is_j^2) \right] e^{-\beta \H(\bm{s})}
\\
 \frac{\partial x_{\sigma_a \sigma_b}}{\partial h} &= \sum_{\bm{s}}  \mathcal{M}_a \mathcal{M}_b 
    \left[ \sum_{i}s_i\right] e^{-\beta \H(\bm{s})} 
\end{align*}
where $\mathcal{M}_{x} \equiv \mathcal{M}(\sigma_x ,s_{i\in x})$ are the cell projection matrices.

When the fixed point is at $L=h=0$ the even and odd couplings decouples and
the stability matrix is block-diagonal, 
with a $3\times 3$ block for even couplings and a $2\times 2$ block for odd
ones.  

The scaling exponents controlling the stability of the fixed point are 
$y_i = \log_b\lambda_i$, where $\lambda_i$ are the eigenvalues of the stability matrix
evaluated at the fixed point, and $b$ the scaling factor of the RG scheme, $b=2$ in this work.

\bigskip

\bibliography{Lucabib}

\begin{thebibliography}{44}
\expandafter\ifx\csname natexlab\endcsname\relax\def\natexlab#1{#1}\fi
\expandafter\ifx\csname bibnamefont\endcsname\relax
  \def\bibnamefont#1{#1}\fi
\expandafter\ifx\csname bibfnamefont\endcsname\relax
  \def\bibfnamefont#1{#1}\fi
\expandafter\ifx\csname citenamefont\endcsname\relax
  \def\citenamefont#1{#1}\fi
\expandafter\ifx\csname url\endcsname\relax
  \def\url#1{\texttt{#1}}\fi
\expandafter\ifx\csname urlprefix\endcsname\relax\def\urlprefix{URL }\fi
\providecommand{\bibinfo}[2]{#2}
\providecommand{\eprint}[2][]{\url{#2}}

\bibitem[{\citenamefont{Kaufman and Griffiths}(1981)}]{Kaufman81}
\bibinfo{author}{\bibfnamefont{M.}~\bibnamefont{Kaufman}} \bibnamefont{and}
  \bibinfo{author}{\bibfnamefont{R.~B.} \bibnamefont{Griffiths}},
  \bibinfo{journal}{Phys. Rev. B} \textbf{\bibinfo{volume}{24}},
  \bibinfo{pages}{496} (\bibinfo{year}{1981}).

\bibitem[{\citenamefont{Griffiths and Kaufman}(1982)}]{Kaufman82}
\bibinfo{author}{\bibfnamefont{R.~B.} \bibnamefont{Griffiths}}
  \bibnamefont{and} \bibinfo{author}{\bibfnamefont{M.}~\bibnamefont{Kaufman}},
  \bibinfo{journal}{Phys. Rev. B} \textbf{\bibinfo{volume}{26}},
  \bibinfo{pages}{5022} (\bibinfo{year}{1982}).

\bibitem[{\citenamefont{Salmon et~al.}(2010)\citenamefont{Salmon, Agostini, and
  Nobre}}]{Salmon2010}
\bibinfo{author}{\bibfnamefont{O.~R.} \bibnamefont{Salmon}},
  \bibinfo{author}{\bibfnamefont{B.~T.} \bibnamefont{Agostini}},
  \bibnamefont{and} \bibinfo{author}{\bibfnamefont{F.~D.} \bibnamefont{Nobre}},
  \bibinfo{journal}{Physics Letters A} \textbf{\bibinfo{volume}{374}},
  \bibinfo{pages}{1631 } (\bibinfo{year}{2010}).

\bibitem[{\citenamefont{Antenucci et~al.}(2014)\citenamefont{Antenucci,
  Crisanti, and Leuzzi}}]{ACL_HL}
\bibinfo{author}{\bibfnamefont{F.}~\bibnamefont{Antenucci}},
  \bibinfo{author}{\bibfnamefont{A.}~\bibnamefont{Crisanti}}, \bibnamefont{and}
  \bibinfo{author}{\bibfnamefont{L.}~\bibnamefont{Leuzzi}},
  \bibinfo{journal}{Journal of Statistical Physics}
  \textbf{\bibinfo{volume}{155}}, \bibinfo{pages}{909} (\bibinfo{year}{2014}),
  ISSN \bibinfo{issn}{0022-4715}.

\bibitem[{\citenamefont{Niemeijer and Leeuwen}(1973)}]{Niemeijer73}
\bibinfo{author}{\bibfnamefont{T.}~\bibnamefont{Niemeijer}} \bibnamefont{and}
  \bibinfo{author}{\bibfnamefont{J.~M.~J.} \bibnamefont{Leeuwen}},
  \bibinfo{journal}{Phys. Rev. Lett.} \textbf{\bibinfo{volume}{31}},
  \bibinfo{pages}{1411} (\bibinfo{year}{1973}).

\bibitem[{\citenamefont{Berker and Wortis}(1976)}]{Berker76}
\bibinfo{author}{\bibfnamefont{A.~N.} \bibnamefont{Berker}} \bibnamefont{and}
  \bibinfo{author}{\bibfnamefont{M.}~\bibnamefont{Wortis}},
  \bibinfo{journal}{Physical Review B} \textbf{\bibinfo{volume}{14}},
  \bibinfo{pages}{4946} (\bibinfo{year}{1976}).

\bibitem[{\citenamefont{Reynolds et~al.}(1977)\citenamefont{Reynolds, Stanley,
  and Klein}}]{Percolation77}
\bibinfo{author}{\bibfnamefont{P.~J.} \bibnamefont{Reynolds}},
  \bibinfo{author}{\bibfnamefont{H.~E.} \bibnamefont{Stanley}},
  \bibnamefont{and} \bibinfo{author}{\bibfnamefont{W.}~\bibnamefont{Klein}},
  \bibinfo{journal}{Journal of Physics C: Solid State Physics}
  \textbf{\bibinfo{volume}{10}}, \bibinfo{pages}{L167} (\bibinfo{year}{1977}).

\bibitem[{\citenamefont{Reynolds et~al.}(1978)\citenamefont{Reynolds, Stanley,
  and Klein}}]{Percolation78}
\bibinfo{author}{\bibfnamefont{P.~J.} \bibnamefont{Reynolds}},
  \bibinfo{author}{\bibfnamefont{H.~E.} \bibnamefont{Stanley}},
  \bibnamefont{and} \bibinfo{author}{\bibfnamefont{W.}~\bibnamefont{Klein}},
  \bibinfo{journal}{Journal of Physics A: Mathematical and General}
  \textbf{\bibinfo{volume}{11}}, \bibinfo{pages}{L199} (\bibinfo{year}{1978}).

\bibitem[{\citenamefont{Nishimori}(2001)}]{Nishimori01}
\bibinfo{author}{\bibfnamefont{H.}~\bibnamefont{Nishimori}},
  \emph{\bibinfo{title}{{Statistical Physics of Spin Glasses and Information
  Processing: An Introduction}}} (\bibinfo{publisher}{Oxford University Press
  (Oxford)}, \bibinfo{year}{2001}).

\bibitem[{\citenamefont{Nienhuis and Nauenberg}(1975)}]{Nienhuis75}
\bibinfo{author}{\bibfnamefont{B.}~\bibnamefont{Nienhuis}} \bibnamefont{and}
  \bibinfo{author}{\bibfnamefont{M.}~\bibnamefont{Nauenberg}},
  \bibinfo{journal}{Phys. Rev. Lett.} \textbf{\bibinfo{volume}{35}},
  \bibinfo{pages}{477} (\bibinfo{year}{1975}).

\bibitem[{\citenamefont{Onsager}(1943)}]{Onsager43}
\bibinfo{author}{\bibfnamefont{L.}~\bibnamefont{Onsager}},
  \bibinfo{journal}{Phys. Rev.} \textbf{\bibinfo{volume}{65}},
  \bibinfo{pages}{117} (\bibinfo{year}{1943}).

\bibitem[{\citenamefont{Southern and Young}(1977)}]{Southern77}
\bibinfo{author}{\bibfnamefont{B.~W.} \bibnamefont{Southern}} \bibnamefont{and}
  \bibinfo{author}{\bibfnamefont{A.~P.} \bibnamefont{Young}},
  \bibinfo{journal}{Journal of Physics C} \textbf{\bibinfo{volume}{10}},
  \bibinfo{pages}{2179} (\bibinfo{year}{1977}).

\bibitem[{\citenamefont{Nobre}(2001)}]{Nobre01}
\bibinfo{author}{\bibfnamefont{F.~D.} \bibnamefont{Nobre}},
  \bibinfo{journal}{Physical Review E} \textbf{\bibinfo{volume}{64}},
  \bibinfo{pages}{046108} (\bibinfo{year}{2001}).

\bibitem[{\citenamefont{Nishimori}(1981)}]{Nishimori81}
\bibinfo{author}{\bibfnamefont{H.}~\bibnamefont{Nishimori}},
  \bibinfo{journal}{Prog. Theor. Phys.} \textbf{\bibinfo{volume}{66}},
  \bibinfo{pages}{1169} (\bibinfo{year}{1981}).

\bibitem[{\citenamefont{Harris}(1974)}]{Harris74}
\bibinfo{author}{\bibfnamefont{A.~B.} \bibnamefont{Harris}},
  \bibinfo{journal}{J. Phys. C: Sol. St. Phys.} \textbf{\bibinfo{volume}{7}},
  \bibinfo{pages}{1671} (\bibinfo{year}{1974}).

\bibitem[{\citenamefont{Toldin et~al.}(2009)\citenamefont{Toldin, Pelissetto,
  and Vicari}}]{Toldin09}
\bibinfo{author}{\bibfnamefont{F.~P.} \bibnamefont{Toldin}},
  \bibinfo{author}{\bibfnamefont{A.}~\bibnamefont{Pelissetto}},
  \bibnamefont{and} \bibinfo{author}{\bibfnamefont{E.}~\bibnamefont{Vicari}},
  \bibinfo{journal}{JSTAT} \textbf{\bibinfo{volume}{135}},
  \bibinfo{pages}{1039} (\bibinfo{year}{2009}).

\bibitem[{\citenamefont{Hasenbusch et~al.}(2008)\citenamefont{Hasenbusch,
  Toldin, Pelissetto, and Vicari}}]{Pelissetto08}
\bibinfo{author}{\bibfnamefont{M.}~\bibnamefont{Hasenbusch}},
  \bibinfo{author}{\bibfnamefont{F.~P.} \bibnamefont{Toldin}},
  \bibinfo{author}{\bibfnamefont{A.}~\bibnamefont{Pelissetto}},
  \bibnamefont{and} \bibinfo{author}{\bibfnamefont{E.}~\bibnamefont{Vicari}},
  \bibinfo{journal}{Phys. Rev. E} \textbf{\bibinfo{volume}{77}},
  \bibinfo{pages}{051115} (\bibinfo{year}{2008}).

\bibitem[{\citenamefont{Ohzeki}(2009)}]{Ohzeki09c}
\bibinfo{author}{\bibfnamefont{M.}~\bibnamefont{Ohzeki}},
  \bibinfo{journal}{Phys. Rev. E} \textbf{\bibinfo{volume}{79}},
  \bibinfo{pages}{021129} (\bibinfo{year}{2009}).

\bibitem[{\citenamefont{Talapov and Bl{\"o}te}(1996)}]{Talapov96}
\bibinfo{author}{\bibfnamefont{A.~L.} \bibnamefont{Talapov}} \bibnamefont{and}
  \bibinfo{author}{\bibfnamefont{H.~W.~J.} \bibnamefont{Bl{\"o}te}},
  \bibinfo{journal}{J. of Phys. A} \textbf{\bibinfo{volume}{29}},
  \bibinfo{pages}{5727} (\bibinfo{year}{1996}).

\bibitem[{\citenamefont{Pelissetto and Vicari}(2002)}]{Pelissetto02}
\bibinfo{author}{\bibfnamefont{A.}~\bibnamefont{Pelissetto}} \bibnamefont{and}
  \bibinfo{author}{\bibfnamefont{E.}~\bibnamefont{Vicari}},
  \bibinfo{journal}{Phys. Rep.} \textbf{\bibinfo{volume}{368}},
  \bibinfo{pages}{549} (\bibinfo{year}{2002}).

\bibitem[{\citenamefont{Ozeki and Ito}(1998)}]{Ozeki98}
\bibinfo{author}{\bibfnamefont{Y.}~\bibnamefont{Ozeki}} \bibnamefont{and}
  \bibinfo{author}{\bibfnamefont{N.}~\bibnamefont{Ito}},
  \bibinfo{journal}{Journal of Physics A: Mathematical and General}
  \textbf{\bibinfo{volume}{31}}, \bibinfo{pages}{5451} (\bibinfo{year}{1998}).

\bibitem[{\citenamefont{Blume et~al.}(1971)\citenamefont{Blume, Emery, and
  Griffiths}}]{Blume71}
\bibinfo{author}{\bibfnamefont{M.}~\bibnamefont{Blume}},
  \bibinfo{author}{\bibfnamefont{V.~J.} \bibnamefont{Emery}}, \bibnamefont{and}
  \bibinfo{author}{\bibfnamefont{R.~B.} \bibnamefont{Griffiths}},
  \bibinfo{journal}{Phys. Rev. A} \textbf{\bibinfo{volume}{4}},
  \bibinfo{pages}{1071} (\bibinfo{year}{1971}).

\bibitem[{\citenamefont{Blume}(1966)}]{Blume66}
\bibinfo{author}{\bibfnamefont{M.}~\bibnamefont{Blume}},
  \bibinfo{journal}{Phys. Rev.} \textbf{\bibinfo{volume}{141}},
  \bibinfo{pages}{517} (\bibinfo{year}{1966}).

\bibitem[{\citenamefont{Capel}(1966)}]{Capel66}
\bibinfo{author}{\bibfnamefont{H.~W.} \bibnamefont{Capel}},
  \bibinfo{journal}{Physica} \textbf{\bibinfo{volume}{32}},
  \bibinfo{pages}{966} (\bibinfo{year}{1966}).

\bibitem[{\citenamefont{Saul et~al.}(1974)\citenamefont{Saul, Wortis, and
  Stauffer}}]{Saul74}
\bibinfo{author}{\bibfnamefont{D.~M.} \bibnamefont{Saul}},
  \bibinfo{author}{\bibfnamefont{M.}~\bibnamefont{Wortis}}, \bibnamefont{and}
  \bibinfo{author}{\bibfnamefont{D.}~\bibnamefont{Stauffer}},
  \bibinfo{journal}{Phys. Rev. B} \textbf{\bibinfo{volume}{9}},
  \bibinfo{pages}{4964} (\bibinfo{year}{1974}).

\bibitem[{\citenamefont{Deserno}(1997)}]{Deserno97}
\bibinfo{author}{\bibfnamefont{M.}~\bibnamefont{Deserno}},
  \bibinfo{journal}{Phys. Rev. E} \textbf{\bibinfo{volume}{56}},
  \bibinfo{pages}{5204} (\bibinfo{year}{1997}).

\bibitem[{\citenamefont{Chakraborty}(1984)}]{Chakraborty84}
\bibinfo{author}{\bibfnamefont{K.~G.} \bibnamefont{Chakraborty}},
  \bibinfo{journal}{Phys. Rev. B} \textbf{\bibinfo{volume}{29}},
  \bibinfo{pages}{1454} (\bibinfo{year}{1984}).

\bibitem[{\citenamefont{Baran and Levitskii}(2002)}]{Baran02}
\bibinfo{author}{\bibfnamefont{O.~R.} \bibnamefont{Baran}} \bibnamefont{and}
  \bibinfo{author}{\bibfnamefont{R.~R.} \bibnamefont{Levitskii}},
  \bibinfo{journal}{Phys. Rev. B} \textbf{\bibinfo{volume}{65}},
  \bibinfo{pages}{172407} (\bibinfo{year}{2002}).

\bibitem[{\citenamefont{Crisanti and Leuzzi}(2002)}]{Crisanti02}
\bibinfo{author}{\bibfnamefont{A.}~\bibnamefont{Crisanti}} \bibnamefont{and}
  \bibinfo{author}{\bibfnamefont{L.}~\bibnamefont{Leuzzi}},
  \bibinfo{journal}{Phys. Rev. Lett.} \textbf{\bibinfo{volume}{89}},
  \bibinfo{pages}{237204} (\bibinfo{year}{2002}).

\bibitem[{\citenamefont{Crisanti and Ritort}(2004)}]{Crisanti04}
\bibinfo{author}{\bibfnamefont{A.}~\bibnamefont{Crisanti}} \bibnamefont{and}
  \bibinfo{author}{\bibfnamefont{F.}~\bibnamefont{Ritort}},
  \bibinfo{journal}{\em Europhys. Lett.\em} \textbf{\bibinfo{volume}{{\bf
  66}}}, \bibinfo{pages}{253} (\bibinfo{year}{2004}).

\bibitem[{\citenamefont{Crisanti et~al.}(2005)\citenamefont{Crisanti, Leuzzi,
  and Rizzo}}]{Crisanti05}
\bibinfo{author}{\bibfnamefont{A.}~\bibnamefont{Crisanti}},
  \bibinfo{author}{\bibfnamefont{L.}~\bibnamefont{Leuzzi}}, \bibnamefont{and}
  \bibinfo{author}{\bibfnamefont{T.}~\bibnamefont{Rizzo}},
  \bibinfo{journal}{\em Phys. Rev. B\em} \textbf{\bibinfo{volume}{{\bf 71}}},
  \bibinfo{pages}{094202} (\bibinfo{year}{2005}).

\bibitem[{\citenamefont{Falicov and Berker}(1996)}]{Falicov96}
\bibinfo{author}{\bibfnamefont{A.}~\bibnamefont{Falicov}} \bibnamefont{and}
  \bibinfo{author}{\bibfnamefont{A.~N.} \bibnamefont{Berker}},
  \bibinfo{journal}{Phys. Rev. Lett.} \textbf{\bibinfo{volume}{76}},
  \bibinfo{pages}{4380} (\bibinfo{year}{1996}).

\bibitem[{\citenamefont{\"Oz\ifmmode~\mbox{\c{c}}\else \c{c}\fi{}elik and
  Berker}(2008)}]{Ozcelik08}
\bibinfo{author}{\bibfnamefont{V.~O.}
  \bibnamefont{\"Oz\ifmmode~\mbox{\c{c}}\else \c{c}\fi{}elik}}
  \bibnamefont{and} \bibinfo{author}{\bibfnamefont{A.~N.}
  \bibnamefont{Berker}}, \bibinfo{journal}{Phys. Rev. E}
  \textbf{\bibinfo{volume}{78}}, \bibinfo{pages}{031104}
  (\bibinfo{year}{2008}).

\bibitem[{\citenamefont{Puha and Diep}({2000})}]{Puha00}
\bibinfo{author}{\bibfnamefont{I.}~\bibnamefont{Puha}} \bibnamefont{and}
  \bibinfo{author}{\bibfnamefont{H.~T.} \bibnamefont{Diep}},
  \bibinfo{journal}{{J. Mag. Mag. Mat.}} \textbf{\bibinfo{volume}{{224}}},
  \bibinfo{pages}{85} (\bibinfo{year}{{2000}}).

\bibitem[{\citenamefont{Paoluzzi et~al.}(2010)\citenamefont{Paoluzzi, Leuzzi,
  and Crisanti}}]{Paoluzzi10}
\bibinfo{author}{\bibfnamefont{M.}~\bibnamefont{Paoluzzi}},
  \bibinfo{author}{\bibfnamefont{L.}~\bibnamefont{Leuzzi}}, \bibnamefont{and}
  \bibinfo{author}{\bibfnamefont{A.}~\bibnamefont{Crisanti}},
  \bibinfo{journal}{Phys. Rev. Lett.} \textbf{\bibinfo{volume}{\bf 104}},
  \bibinfo{pages}{120602} (\bibinfo{year}{2010}).

\bibitem[{\citenamefont{Paoluzzi et~al.}(2011)\citenamefont{Paoluzzi, Leuzzi,
  and Crisanti}}]{Paoluzzi11}
\bibinfo{author}{\bibfnamefont{M.}~\bibnamefont{Paoluzzi}},
  \bibinfo{author}{\bibfnamefont{L.}~\bibnamefont{Leuzzi}}, \bibnamefont{and}
  \bibinfo{author}{\bibfnamefont{A.}~\bibnamefont{Crisanti}},
  \bibinfo{journal}{Philos. Mag.} \textbf{\bibinfo{volume}{91}},
  \bibinfo{pages}{1966} (\bibinfo{year}{2011}).

\bibitem[{\citenamefont{Leuzzi et~al.}(2011)\citenamefont{Leuzzi, Paoluzzi, and
  Crisanti}}]{Leuzzi11}
\bibinfo{author}{\bibfnamefont{L.}~\bibnamefont{Leuzzi}},
  \bibinfo{author}{\bibfnamefont{M.}~\bibnamefont{Paoluzzi}}, \bibnamefont{and}
  \bibinfo{author}{\bibfnamefont{A.}~\bibnamefont{Crisanti}},
  \bibinfo{journal}{Phys. Rev. B} \textbf{\bibinfo{volume}{\bf 83}},
  \bibinfo{pages}{014107} (\bibinfo{year}{2011}).

\bibitem[{\citenamefont{Griffiths}(1967)}]{Griffiths}
\bibinfo{author}{\bibfnamefont{R.}~\bibnamefont{Griffiths}},
  \bibinfo{journal}{Physica} \textbf{\bibinfo{volume}{33}}, \bibinfo{pages}{689
  } (\bibinfo{year}{1967}).

\bibitem[{\citenamefont{Wu}(1982)}]{PottsRev}
\bibinfo{author}{\bibfnamefont{F.~Y.} \bibnamefont{Wu}}, \bibinfo{journal}{Rev.
  Mod. Phys.} \textbf{\bibinfo{volume}{54}}, \bibinfo{pages}{235}
  (\bibinfo{year}{1982}).

\bibitem[{\citenamefont{Kinzel and Fisher}(1978)}]{Kinzel78}
\bibinfo{author}{\bibfnamefont{W.}~\bibnamefont{Kinzel}} \bibnamefont{and}
  \bibinfo{author}{\bibfnamefont{F.~H.} \bibnamefont{Fisher}},
  \bibinfo{journal}{J. Phys. C} \textbf{\bibinfo{volume}{11}},
  \bibinfo{pages}{2115} (\bibinfo{year}{1978}).

\bibitem[{\citenamefont{Tatsumi}(1978)}]{Tatsumi78}
\bibinfo{author}{\bibfnamefont{T.}~\bibnamefont{Tatsumi}},
  \bibinfo{journal}{Prog. Theor. Phys.} \textbf{\bibinfo{volume}{59}},
  \bibinfo{pages}{405} (\bibinfo{year}{1978}).

\bibitem[{\citenamefont{Franz et~al.}(1994)\citenamefont{Franz, Parisi, and
  Virasoro}}]{Franz94}
\bibinfo{author}{\bibfnamefont{S.}~\bibnamefont{Franz}},
  \bibinfo{author}{\bibfnamefont{G.}~\bibnamefont{Parisi}}, \bibnamefont{and}
  \bibinfo{author}{\bibfnamefont{M.-A.} \bibnamefont{Virasoro}},
  \bibinfo{journal}{J. Phys. I (France)} \textbf{\bibinfo{volume}{4}},
  \bibinfo{pages}{1657} (\bibinfo{year}{1994}).

\bibitem[{\citenamefont{Franz and Toninelli}(2005)}]{Franz05b}
\bibinfo{author}{\bibfnamefont{S.}~\bibnamefont{Franz}} \bibnamefont{and}
  \bibinfo{author}{\bibfnamefont{F.}~\bibnamefont{Toninelli}},
  \bibinfo{journal}{J. Stat. Mech.} p. \bibinfo{pages}{P01008}
  (\bibinfo{year}{2005}).

\bibitem[{\citenamefont{Boettcher}(2005)}]{Boettcher05}
\bibinfo{author}{\bibfnamefont{S.}~\bibnamefont{Boettcher}},
  \bibinfo{journal}{Phys. Rev. Lett.} \textbf{\bibinfo{volume}{95}},
  \bibinfo{pages}{197205} (\bibinfo{year}{2005}).

\end{thebibliography}
\end{document}